# Local energy landscape drives long exciton diffusion in 2D halide perovskite semiconductors

*Alan Baldwin [a,b §], Géraud Delport [a, §,†], Kai Leng [c,d], Rosemonde Chahbazian [a], Krzysztof Galkowski [a,e], Kian Ping Loh [d] and Samuel D. Stranks [a,b]\**

a) Cavendish Laboratory, University of Cambridge, JJ Thomson Avenue, Cambridge CB3 0HE, UK.

b) Department of Chemical Engineering & Biotechnology, University of Cambridge, Philippa Fawcett Drive, Cambridge CB3 0AS, UK

c) Department of Applied Physics, The Hong Kong Polytechnic University, Hung Hom, Kowloon, Hong Kong, China.

d) Department of Chemistry, National University of Singapore, Singapore, Singapore.

e) Institute of Physics, Faculty of Physics, Astronomy and Informatics, Nicolaus Copernicus University, 5th Grudziadzka St., 87−100 Toruń, Poland

KEYWORDS. Halide Perovskites, 2D semiconductors, exciton diffusion, exciton-phonon coupling, trapping dynamics.

Halide perovskites have emerged as disruptive semiconductors for applications including photovoltaics and light emitting devices, with modular optoelectronic properties realisable through composition and dimensionality tuning. Layered Ruddlesden-Popper perovskites of the form $BA_2MA_{n-1}Pb_nI_{3n+1}$, where *n* is the number of lead-halide and methylammonium (MA) sheets spaced by longer butylammonium (BA) cations, are particularly interesting due to their unique two-dimensional character and charge carrier dynamics dominated by strongly bound excitons. However, long-range energy transport through exciton diffusion in these materials is not understood or realised. Here, we employ local time-resolved luminescence mapping techniques to visualise exciton transport in high-quality exfoliated flakes of the $BA_2MA_{n-1}Pb_nI_{3n+1}$ perovskite family. We uncover two distinct transport regimes, depending on the temperature range studied. At temperatures above 100 K, diffusion is mediated by thermally activated hopping processes between localised states. At lower temperatures, a non-uniform energetic landscape emerges in which exciton transport is dominated by energy funnelling processes to lower energy states, leading to long range transport over hundreds of nanometres even in the absence of exciton-phonon coupling and in the presence of local optoelectronic heterogeneity. Efficient, long-range and switchable excitonic funnelling offers exciting possibilities of controlled directional long-range transport in these 2D materials for new device applications.

Halide perovskites have intrigued the semiconductor community for the last decade, challenging the established understanding that defect concentrations in semiconductors must be minimised at all costs for high device performance[1], which typically requires complex fabrication methods. Remarkably, the perovskite structure offers the possibility to tune the crystal dimensionality and the inherent photophysical properties, for example by changing the length and nature of the A-site cation out of a large library of compatible molecules[2]. Ruddlesden-Popper perovskites (RPPs) hold a special place in the perovskite family due to their unique two dimensional character (Figure 1a) and their enhanced stability with respect to conventional 3D perovskites to atmosphere and under illumination[3,4]. These materials take the general formula $S_2A_{n-1}B_nX_{3n+1}$ where S represents a large organic spacer cation, A a small monovalent cation, B a divalent metal and X a halide anion.

Structurally, they are composed of corner-sharing $BX_6$ octahedra quasi-2D layers, with *n* representing the number of $BX_6$ planes, with the layers separated by bulky S molecules[5–7]. The two-dimensionality of each of the layers coupled with the low dielectric constant of the spacer molecules results in charge dynamics which are dominated by strongly bound excitons[5,8–10]. The remarkable photophysical properties of these RPPs have been utilised in various domains such as lasers[11], LEDs[12] and in efficient, stable perovskite solar cells[13]. High-quality single crystal RPPs can be fabricated and mechanically exfoliated to yield any desired number of layers[6,14–16], paving the way for ultrathin optoelectronic devices[17]. Recent macroscopic reports have focused on the exciton dynamics[18] in such materials, including nonlinear[19–21] properties such as exciton-exciton annihilation[7,22] and the influence of the size of the quantum well and the employed organic spacer[7,22–27]. However, the fundamental processes related to the transport of excitons in such systems is under hot debate[22,26,28,29]. Many aspects of exciton recombination and transport, particularly on the local scale, are not understood but such understanding will be critical for ultimately attaining fine control over energy transport in device applications.

Here, we study exciton dynamics in pure phase $BA_2MA_{n-1}Pb_nI_{3n+1}$ (BA = butylammonium, MA = methylammonium) flakes focusing on the *n* = 4 system, with similar results obtained for n = 2, exfoliated from macroscopic single crystals (see SI Section I for preparation details). We employ time-resolved photoluminescence microscopy (TRPL)[30–34], coupled to a helium cryostat, to visualise the local diffusion properties as a function of temperature (see SI Sections II, III and Figure S1 for a description of the customised confocal photoluminescence setup). We show that at temperatures 300 – 100 K, exciton diffusion is mediated by thermally activated hopping through localised states that are also responsible for the non-radiative decay of these excitons. At temperatures below 100 K, such hopping processes are inefficient and we instead observe a transport regime in which exciton transport is driven by energy gradients across a varying bandgap landscape originating from local phase or structural variations. Such energy gradients lead to exciton transport on length scales of ≈ 500 nm associated with strong luminescence from low energy regions of the crystal. This study reveals that long-range directional energy transport is possible if local energetic landscapes can be controlled and exploited, paving the way towards new devices based on 2D materials including transistors, light emitting devices and photodetectors.

We focus on the exciton dynamics in $n = 4$ exfoliated thin RPP crystals deposited on a silicon substrate and encapsulated with a hexagonal Boron-Nitride (h-BN) layer to ensure environmental and illumination stability[17] (see SI Section I for details and Figure 1a for schematic structure of $n = 4$ sample). We also observe similar results to those reported here for $n = 2$ crystals (SI Section X). An optical image of one of the flakes of interest is shown in Figure 1b with thickness of ≈9 stacked quantum wells (≈30 nm thickness) based on the optical contrast of the flake and on previous studies[35] (see SI Section V and Figure S2 for details). Such thin flakes maximize our chances to accurately separate the exciton diffusion process from other physical effects to accurately capture the thermal evolution of the excitonic diffusion and recombination process. Specifically, photonic effects (reabsorption, recycling or lateral waveguiding effects[36–38]) will be negligible in both the vertical and lateral directions and phonon-mediated phases transitions are forbidden in such thin flakes[39].

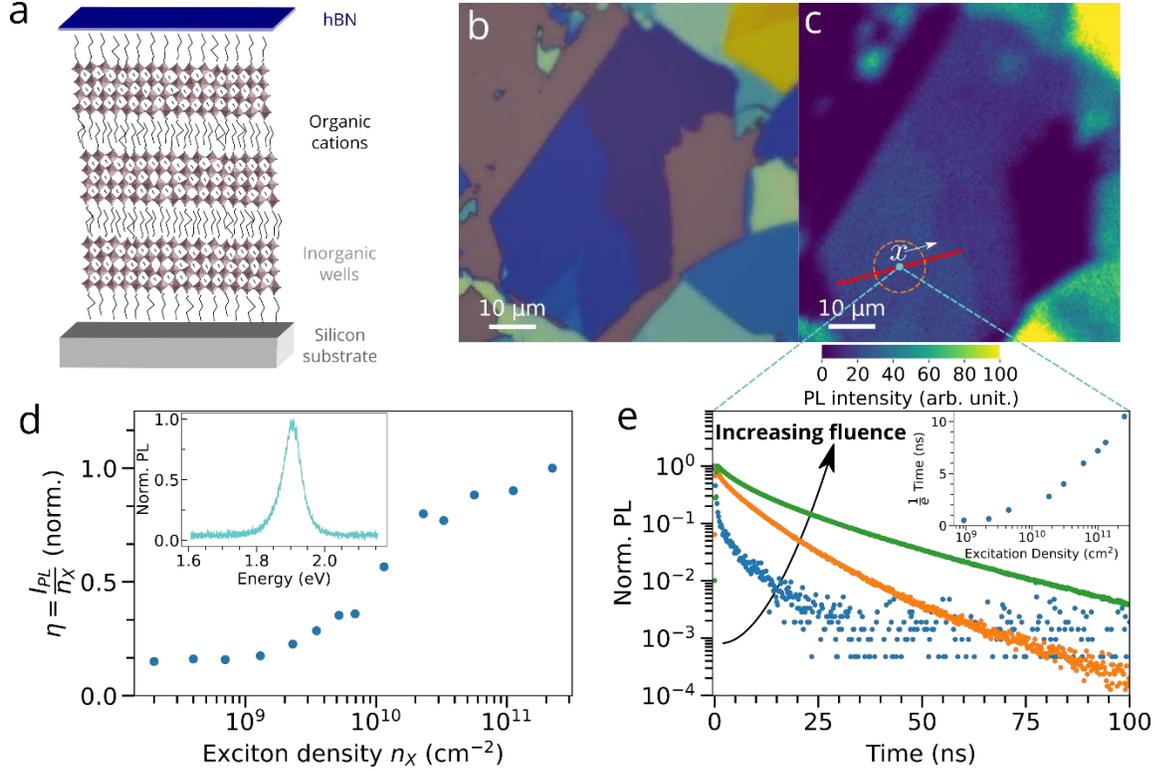

**Figure 1: Local photoluminescence properties of the studied BA$_2$MA$_3$Pb$_4$I$_{13}$ $n$=4 RRP flake.** (a) Schematic representation of a few layers $n = 4$ RRP flake. (b) Optical reflection and (c) PL intensity (excitation at 510 nm) images. The red strip in c highlights the region of interest over which transport is subsequently analysed, with the spatial parameter $x$ defined spanning from the centre of the region. (d) Evolution of the normalised PL efficiency η as a function of injected exciton density, $n_X$, with η defined as the ratio of the number of emitted photons (PL intensity, $I_{PL}$) to the number of injected excitons $n_X$, arbitrarily normalised to 1 for clarity. (e) Time-resolved PL decays for increasing injected exciton densities at the centre of the region of interest highlighted on c, with exciton density increasing between $9.2 \times 10^8$ cm$^{-2}$ (blue), $3.0 \times 10^{10}$ cm$^{-2}$ (orange) and $2.5 \times 10^{11}$ cm$^{-2}$ (green). Inset, the normalised PL spectrum at the same location. Measurements were carried out at room temperature.

Figure 1c displays the photoluminescence (PL) intensity maps of the flake of interest. The local PL spectrum of this flake (Figure 1e inset and Figures S3 and S4) is centred at ≈1.9 eV[18], as expected for a high-quality pure phase $n = 4$ flake, suggesting no other $n$ phases are present in that region. We first measure the local PL properties in the region of interest (Figure 1c) at different excitation regimes with a 510-nm pulsed laser diode (see SI Section VI for details and Figures

S3- 4 for further data). In Figure 1d, we show the PL intensity as a function of excitation density $n_x$, weighted by the excitation density $n_x$, to provide a relative PL efficiency, η (i.e. normalised number of emitted photons per injected exciton; see SI Section VII for further data and Figure S5 for absolute PL intensity). We observe that this curve forms an S-shape comprised of two plateaux separated by a growth phase, which can be ascribed to different regimes of trap state filling[40]. Below $n_x \approx 2 \times 10^9$ cm$^{-2}$ the relative PL efficiency remains constant, indicating that the trap filling process is inefficient in this range of exciton densities. This is explained by the fact that the concentration of available traps remains much larger than that of injected excitons across this excitation regime. Above this $n_x$ value, η rises significantly as the PL intensity increases superlinearly with respect to the exciton density. Across this regime, the exciton density is large enough to fill a non-negligible proportion of traps[7], leading to an increase of the PL efficiency. This trap filling occurs up to a saturation of this effect at $n_x \approx 3 \times 10^{10}$ cm$^{-2}$, above which η again remains constant with increasing excitation density and a large proportion of excitons are recombining radiatively. Similar observations can be made when considering time-resolved PL (TRPL) decays as a function of the exciton density (Figures 1e and S6). In the low exciton density regime (9.2 × 10$^8$ cm$^{-2}$, blue data) corresponding to the first plateau, the TRPL curve exhibits a rapid decay, with a PL lifetime of 0.5 ns defined as the time to fall to $\frac{1}{e}$ of the initial intensity[41], consistent with fast non-radiative decay of trapped excitons as the primary recombination channel. Above this first plateau, the TRPL lifetime increases with increasing excitation density (3.0 × 10$^{10}$ cm$^{-2}$, orange data), until the second plateau is reached (Figures 1e and S7). At this second plateau (2.5 × 10$^{11}$ cm$^{-2}$, green data), we observe a quasi mono-exponential decay with a $\frac{1}{e}$ PL lifetime of 11 ns, which is consistent with classical first order exciton recombination kinetics dominated by a radiative rate that is outcompeting the non-radiative rate.[7]

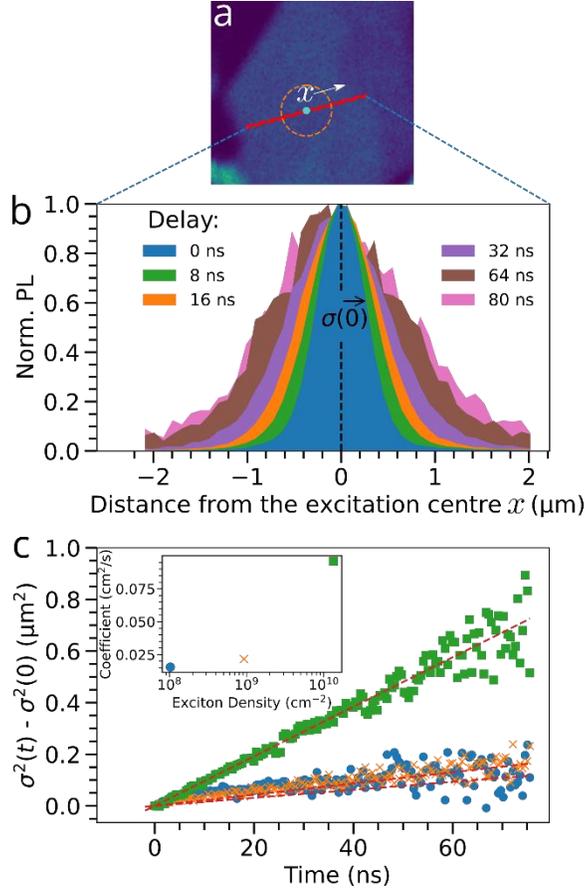

**Figure 2: Local spatially resolved diffusion measurements in an BA$_2$MA$_3$Pb$_4$I$_{13}$ ($n$ = 4) RRP flake at different fluences.** (a) PL map highlighting the region of interest. Scale bar is as per Figure 1c. (b) Selected spatial PL profiles, normalised, at different times after excitation by the laser pulse, localised at $x$ = 0 (centre of the red line in Figure 2a). (c) Spreading profiles showing the temporal evolution of the squared broadening quantities $\sigma^2(t) - \sigma^2(0)$ of the spatial profiles for the flake at room temperature on the region of interest at three different densities of injected excitons. Inset: Diffusion coefficients extracted from fits to the data with Equation 1. To clearly visualise the spreading effect, the profiles shown in (b) were obtained with a 0.8 NA 100 x objective lens (generating a narrower initial exciton distribution), whereas all other data throughout, including the in panel c, were taken with a 0.4 NA 10x objective lens which is more suitable for the cryogenic measurements and does not alter the diffusion measurements (see SI Section VIII and Figure S8).

We now use the same TRPL microscope setup to investigate the impact of this trap filling process on exciton transport. In this configuration, the sample is excited in the same local region at time $t$ = 0 at a fixed position ($x$ = 0) with a Gaussian shaped laser pulse (cf. orange circle Figure 1c), with the PL subsequently collected as a function of time (*t*) at different spatial points (*x*) away from the excitation spot (see Figure S1 for setup). The resulting spectrally integrated spatial PL profiles, from the region highlighted in Figure 2a, are displayed as a function of time in Figure 2b in the low excitation density regime ($n_x \approx 1.0 \times 10^8$ cm$^{-2}$), with the increase in the standard deviation of the Gaussian spread *σ(t)* with time consistent with excited species spreading laterally after excitation (see SI Section IX and Figure S9 for Gaussian profiles at selected times). We find a linear relationship between the quantity [$\sigma^2(t) - \sigma^2(0)$] and *t* (Figure 2c), indicating that the spatial broadening is due to classical exciton diffusion[41,42], and can be related to the exciton diffusion coefficient *D* using the formula [32,42,43]:

$$\sigma^2(t) - \sigma^2(0) = 2Dt \qquad (1).$$

Linear fits to the data in Figure 2c with Equation (1) yields a diffusion coefficient of 0.018 cm$^{-2}$s$^{-1}$ for $n_x = 1.0 \times 10^8$ cm$^{-2}$ (Figure 2c, inset). At this low exciton density, we expect exciton trapping to dominate based on the earlier PL efficiencies, η. At higher exciton densities corresponding to the beginning of the trap saturation regime ($n_x = 9.2 \times 10^8$ cm$^{-2}$), the fitted diffusion coefficient increased slightly to 0.021 cm$^{-2}$s$^{-1}$, and then increases significantly to 0.096 cm$^{-2}$s$^{-1}$ at even higher exciton densities ($n_x = 1.3 \times 10^{10}$ cm$^{-2}$) in which a large proportion of the traps are saturated. We note that there is no significant deviation from the linear evolution of [$\sigma^2(t) - \sigma^2(0)$] across these excitation densities, suggesting that non-linear effects such as exciton-exciton annihilation are not influencing the exciton dynamics. These measurements reveal that the local trap states not only limit the PL efficiency but also limit exciton diffusion in these RPPs systems, consistent with other 2D semiconductors.[26,44] Similar behaviour is observed for an *n* = 2 sample (see SI Section X and Figures S10-13).

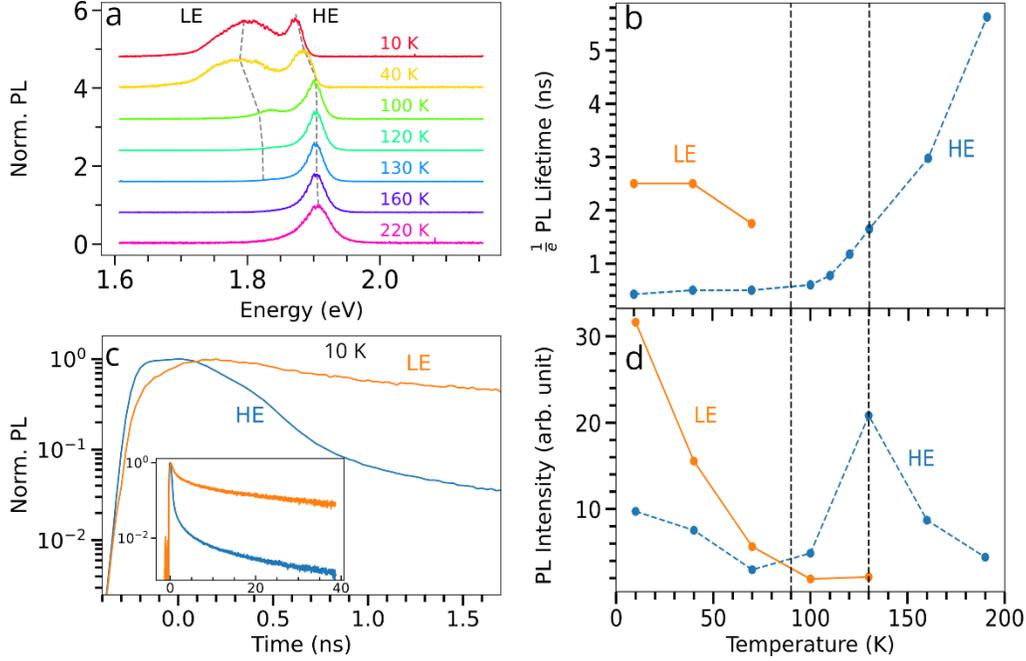

**Figure 3: Temperature-dependent PL spectra and exciton dynamics in a BA$_2$MA$_3$Pb$_4$I$_{13}$ RRP flake,** all measured at the centre of region of interest mentioned earlier. (a) PL spectra showing the evolution of the high energy (HE) peak and the emergence of the low energy (LE) peak at low temperatures, with excitation at 510 nm. Spectra are normalised to the maximum value and vertically offset for clarity. Grey dotted lines tracking the peak centres are guides to the eye. Spectrally selected (b) $\frac{1}{e}$ lifetimes of the LE and high energy (HE) components. (c) Spectrally selected TRPL decays corresponding to the HE and the LE resonance at 10 K, exhibiting the population transfer process. Inset, corresponding decay on the longer time scale. (d) Spectrally selected PL intensities as a function of temperature. Measurements were performed with an excitation density of 9.2 × 10$^8$ cm$^{-2}$. Spectral selection is achieved using appropriate bandpass filters (see SI Section XII and Figure S18).

To further understand the exciton dynamics and transport, we performed temperature-dependent studies on the same region of the *n* = 4 flake (see SI Section XI for further data). We show local PL spectra in Figure 3a as a function of temperature (see Figure S14 for spectra at other temperatures). The position of the main PL resonance, which we refer to herein as the high-energy (HE) peak, does not shift significantly from ≈ 1.9 eV upon decreasing temperature from 300 K to 80 K. This lack of shift is consistent with a strong influence of phonons on the exciton properties[9,45,46] that counterbalance the usual PL redshift in 3D perovskites attributed to the thermal contraction of the crystal within this temperature range[47,48]. The redshift behaviour is recovered over the range 120 to 10 K where the contribution of optical phonons should be negligible (Figure S10). Indeed, the lowest optical phonon mode energy in similar systems have been reported to be ≈10 meV [49,50 51], corresponding to threshold temperatures around ≈120 K. Further evidence for the influence of phonons on the exciton properties is seen in the decrease of the PL resonance width of the HE peak from ≈99 meV at 300 K to 22 meV at 100 K (Figure S15) [45,47].

To further our understanding of the exciton dynamics, we now analyse the evolution of the TRPL decays as a function of temperature (see Figure S16 for all decays). Over the previously studied temperature range, the $\frac{1}{e}$ PL lifetime in the HE peak reduces significantly from 14 ns at 300 K to 6 ns at 200 K and 2.9 ns at 100 K with an exciton density of $9.2 \times 10^8$ cm$^{-2}$ (Figure 3b). Such a general decrease in PL lifetime with decreasing temperature can again, in part, be attributed to a freezing out of phonons, either by the lack of exciton-phonon coupling at low temperature preventing excitons from escaping local traps states or, by analogy with MoSe$_2$[52] or GaAs quantum wells[53], by a lack of phonons that otherwise push excitons away from the bottom of the band and thus prevent recombination until they return. Interestingly, this decrease in PL lifetime is first accompanied by an increase in PL intensity down to 130 K (Figure 3d), but below 130 K the PL intensity reduces significantly. This drop in PL intensity at ≈130 K occurs concomitant with the emergence (Figure 3a) and rapid increase in intensity (Figure 3d) of a second, broader PL resonance at ≈1.8 eV, referred to hereafter as the low-energy (LE) resonance. This LE resonance indicates the presence of radiative states situated energetically below the classical excited state continuum. Such LE resonances could be due to the presence of local domains or traps resulting from the presence of static disorder in the crystal structure[54,55], or impurity phases that are only readily apparent at low temperature[27,54,56,57]. We note that the nature of these localised radiative states appears to be distinct from the non-radiative trap states discussed above, consistent with a general rise of the total PL intensity (LE + HE) observed between 200 K and 10 K (Figure 3d and Figure S17).

The decrease in intensity of the HE peak together with the rise in intensity of the LE peak at low temperature hints at a transfer of exciton population from the HE to LE states[18]. This transfer effect is evident by considering the spectrally resolved PL dynamics (see SI Section XII and Figure S18 for details), at a given temperature, exemplified in Figure 3b at 10 K (see Figure S19 for other temperatures). Within the first nanosecond after the excitation pulse, we observe both a fast decline of the HE component and a rise of the LE component with a rise time of ≈ 0.5 ns (time between the maximum of the HE component and the maximum of the LE component), longer than the setup instrument response of ≈385 ps (Figure S20). After another few hundred picoseconds, the LE component starts declining but at a reduced rate compared to the HE signal. The PL decay of the LE component extends over ≈40 ns (Figure 3c), suggesting that once the excitons reach these LE states, their recombination is slower than on the HE states.

We now consider the temperature-dependent transport properties, and the impact of this energy transfer process on these properties. We start by analysing the diffusion properties associated with the HE peak as a function of temperature at a low exciton density of $n_x = 9.2 \times 10^8$ cm$^{-2}$. We limit our measurements to temperatures below 200 K to maintain a low pressure without disturbing the measurements with the vibrating vacuum cryopump (see SI Sections II-IV). We find that the diffusion coefficient decreases significantly from 0.03 cm$^2$.s$^{-1}$ at 200 K to a negligible value (< 0.001 cm$^2$.s$^{-1}$) below 100 K (see Figures 4c and example fit at 160 K in Figure 4b; see SI Section XV and Figures S21-24 for Gaussian profiles and diffusion fits at different temperatures). This suggests that the excitons became progressively more localised on the HE sites with decreasing temperature.

To consider the transport associated with the LE and HE sites, we show in Figure 4d a PL map of the region of interest at 10 K, showing a spatial distribution of these LE and HE sites. Fluence-dependent measurements show that the emission from the LE states saturate when the excitation density is of the order of $\approx 9 \times 10^8$ cm$^{-2}$ (Figure S25), indicating that a significant portion of the created excitons are unable to access these LE states likely due to a lower concentration of LE regions across the flake, together with slower recombination, compared to the HE regions. This result provides a picture that the spatial energy landscape is comprised of many HE-only domains, sparsely interspersed with LE subdomains. As these LE and HE states are dispersed across the sample, we now consider whether such an exciton transfer system observed can promote a lateral motion of the excitons beyond the classical diffusive process. We show the extracted exciton transport coefficient corresponding to the LE and HE peaks in Figure 4c, again at a low exciton density of $9.2 \times 10^8$ cm$^{-2}$ (see Figure 4a for example from the LE peak at 10 K). A very slow spatial spreading of the PL profile is observed for the HE resonance (Figure S19), comparable to the diffusion coefficient measured just above 100 K. However, the LE band exhibits an initial fast spreading of the spatial PL profile, on the order of a nanosecond, followed by a regime over the next $\approx 80$ ns during which negligible spreading occurs. Further fluence-dependent measurements (Figure 4e) reveal that the transport contribution associated with the LE bands saturates at higher exciton density, consistent with the saturation of the PL from these states. These observations indicate that an efficient lateral motion of the excitons occurs with the same nanosecond timescale as the HE-to-LE population transfer discussed earlier. We propose that the two effects are connected and that the large energy gradient between adjacent HE and LE regions causes excitons to rapidly drift toward the LE states.

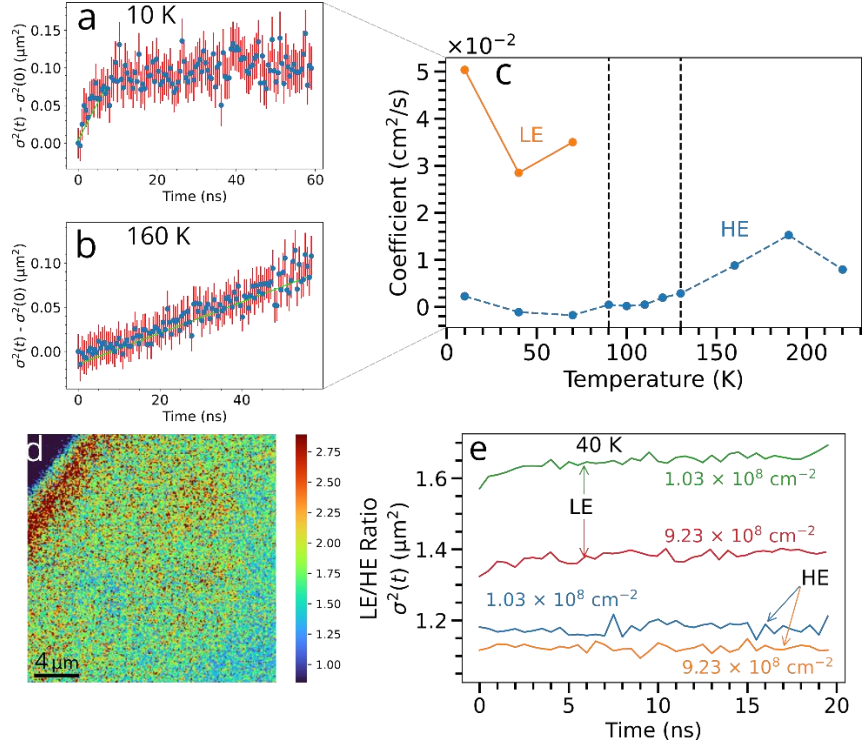

**Figure 4: Temperature-dependent transport properties in a BA$_2$MA$_3$Pb$_4$I$_{13}$ RRP flake.** Spreading profiles at (a) 10 K and (b) 160 K with an excitation density of $9.2 \times 10^8$ cm$^{-2}$. (c) Effective spreading coefficients extracted from fitting the underlying spreading profiles to Equation 1, exemplified by the linear fits in green in a and b, Figures S16 and S19. For the LE components only the initial fast spreading coefficients are shown. (d) Ratio of normalised confocal PL maps at 10 K with excitation at 510 nm of the flake allowing the distinction of the regions containing LE and HE peaks, obtained using bandpass filters. (e) Spreading profiles at 40 K for different injected exciton densities with LE and HE peaks distinguished using the bandpass filters (see SI Section XII and Figure S18).

To validate these interpretations, we model the impact of the LE and HE domains on the exciton transport in this low temperature regime. This simulation uses a finite element approach to solve the corresponding drift-diffusion equation to describe how the local exciton density $n(\vec{r}, t)$ evolves in time $t$ and space $\vec{r}$

$$\frac{\delta n(\vec{r},t)}{\delta t} = -kn(\vec{r},t) + D\nabla^2 n(\vec{r},t) + \mu\nabla\big(n(\vec{r},t)\nabla u(\vec{r})\big) \qquad (2),$$

where the first term on the right hand side corresponds to exciton recombination, $k$, and the second term to exciton diffusion. The third term accounts for the drift of excitons inside the energy gradient, with µ being the exciton mobility and $u(\vec{r})$ the local energetic potential felt by an exciton located at position $\vec{r}$. We fixed the value of several parameters to our experimental results at 10 K (see SI Section XVII for details): specifically, we use distinct exciton recombination rates extracted from TRPL decays (Figure 5d) for the HE and the LE domains of $k_{HE}$ = 0.6 ns$^{-1}$ and $k_{LE}$ = 0.01 ns$^{-1}$, with $k = k_{HE} + k_{LE}$. Furthermore, we model the energy landscape of the sample at 10 K as being composed of adjacent domains of both HE (energy range 1.85 – 1.91 eV) and LE (broader energy range of 1.60 – 1.85 eV) using the spatial distribution from the PL maps in Figure 4a and the PL spectra in Figure 3a. Thus, $D$ is the only free parameter for adjustment. The finite difference mesh employed to solve the equation consists of a 256 x 256 grid, resulting in individual domain sizes equivalent to 80 nm. Using these conditions, we simulated the spatiotemporal evolution of the excitons at 10 K (Figures b-c) and successfully reproduce the main features of the dynamics, in particular the fast spreading of the exciton distribution with spreading coefficient $D$ = 0.69 cm$^2$.s$^{-1}$ closely matching the experimental value of 0.66 cm$^2$.s$^{-1}$ (Figure 5b; see Figure S26 for other simulations). The results from this model validates our understanding of the observations in which the local energy gradient drives the excitons to move downward in energy, which results in a lateral movement of excitons from the HE to the LE domains. The fact that LE sites can saturate allows exploration of further LE sites, leading to long-distance energy transport (≈500 nm) through this exciton funnelling effect.

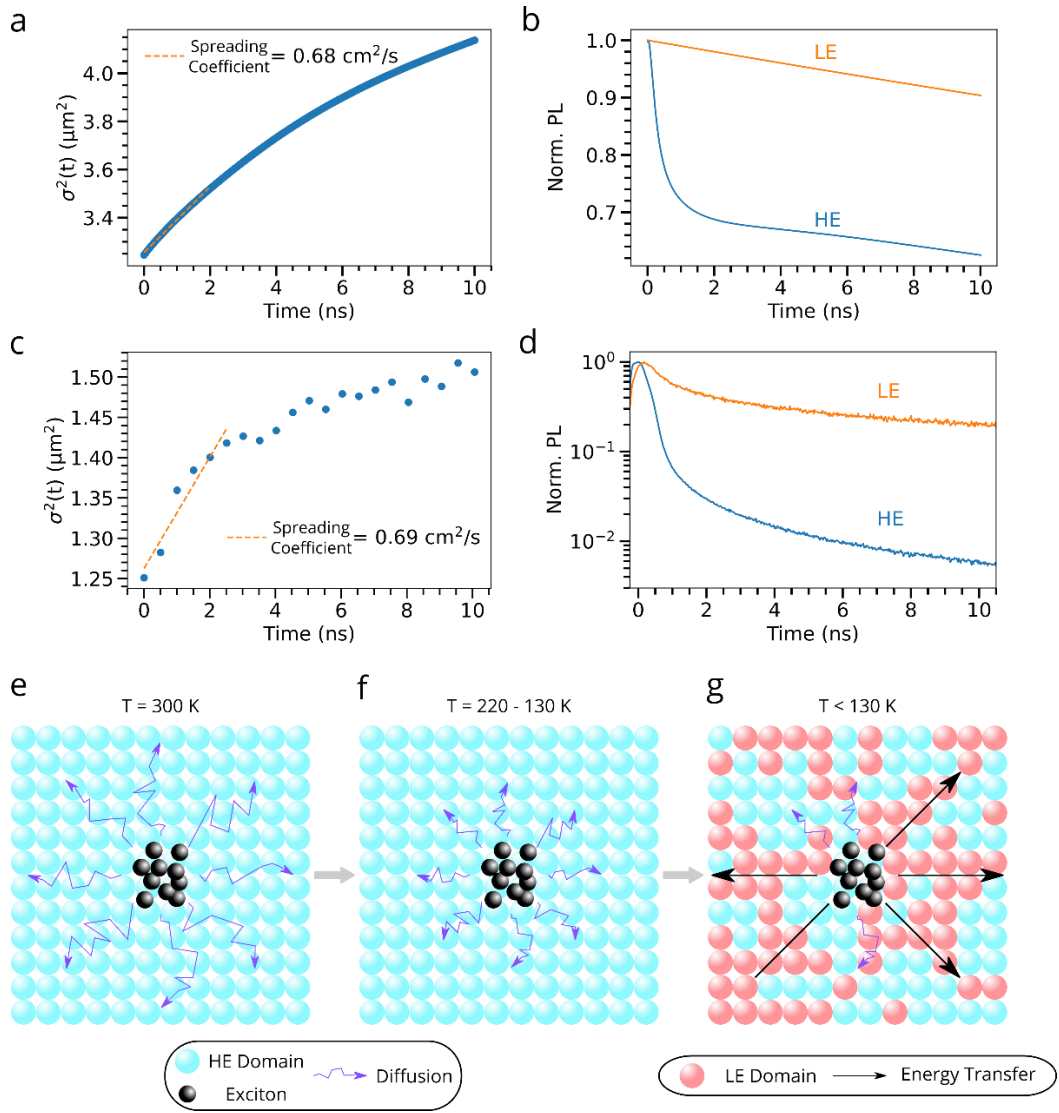

**Figure 5: Understanding the exciton funnelling in a $BA_2MA_3Pb_4I_{13}$ RRP flake at low temperature.** Outputs from simulations of our proposed model, Equation 2, compared to experiment, see SI Section XVII for parameters used. Modelled (a) spreading curve and (b) TRPL decays. Experimental (c) spreading curve and (d) TRPL decays from the actual flake at an exciton density of $9.2 \times 10^8$ cm$^{-2}$. (e) – (g) Schematics to represent the proposed transport model over the studied temperature ranges. (e) At room temperature diffusion is the dominant exciton transport mechanism. (f) As the temperature is lowered the diffusion process becomes less efficient. (g) At T < 130 K, the LE states emerge and energy transfer becomes the dominant transport mechanism. The lengths of the diffusion and energy transfer indicators are indicative of the rates of these processes.

Based on our collective results, we propose a general description of the exciton dynamics in these RPP flakes that is summarised in Figures 5e-g. At room temperature and down to ≈100 K, exciton transport is governed by an interplay between non-radiative trap states and phonons. In this picture (Figure 5e), exciton diffusion occurs via a hopping process in which significant thermal energy provided by phonons is required to promote excitons from localised states toward the excited state continuum. Such a diffusion process is efficient at high temperatures, though this also means that excitons can more readily find deep trap states on which they recombine non-radiatively; hence the PL efficiency generally increases with decreasing temperature. Diffusion becomes less efficient at lower temperature as fewer phonons states are populated (Figure 5f) and, at or below 100 K, the exciton-phonon coupling effects freeze out due to a negligible population of optical phonons[9,49,50] in such RPP flakes. At ≈ 130 K, a series of broad, lower energy (LE) states are detectable. At temperatures below 100 K, excitons are split into two populations: one which is localised before recombining, leading to the HE resonance. In parallel, a fraction of these excitons move laterally as they funnel down into the LE sites (Figure 5g). This funnelling process recovers long-range exciton transport (≈500 nm) at low temperature even in spite of the absence of phonon population. The long-range transport seems to be uniquely enabled by the ability to saturate local LE sites at sufficiently high exciton density. If exploited efficiently, for example through controlled design of the spatial distribution of these LE states, this can lead to spatial control of exciton motion over even larger length scales. Furthermore, the fact the LE sites can saturate implies that their properties can be turned on and off at will, for instance by modulating the exciton density or by applying an external stress (mechanical, illumination[17]) that would influence the static disorder. If properly controlled, such a property could be used to tailor the exciton properties (mobility, luminescence) in a device towards a given application requirement at a given time.

In conclusion, we studied the local temperature-dependent dynamics and transport of excitons in thin Ruddlesden-Popper perovskite flakes using time-resolved luminescence microscopy. We identified the existence of several temperature regimes and highlighted the interplay between excitons, phonons and the local energetic landscape. Below 100 K, the exciton diffusion process is inefficient due to the absence of optical phonons. Yet, we showed that excitons can still travel through a static disorder landscape scheme in which long-lived radiative shallow states play an important role. At higher temperatures, the progressive population of the phonon modes causes dynamic disorder to dominate over static disorder. In parallel, the exciton-phonon coupling allows the exciton to move across the crystal in a diffusive fashion, hopping from one shallow trap to the next until they either recombine radiatively or they encounter non-radiative deep trap states. Phonons induce dynamic disorder, increasing the exciton mobility, enabling the excitons to reach non-radiative deep traps. This study provides new microscopic visualisation of the fundamental mechanisms underlying the exciton dynamics in these RPP flakes, and highlights the major role played by the energetic landscape in the diffusion process in such 2D materials. With fine control over such a landscape, and in particular the presence of non-radiative deep traps states, we could control the efficiency of the excitonic motion and design highly efficient optoelectronics devices based on these Ruddlesden-Popper perovskites.

ASSOCIATED CONTENT

**Supporting Information**.

The Supporting Information is available free of charge at

Experimental section, including the synthesis of the 2D perovskites crystal, and the evaluation of the thickness of the flakes, details of the spectral PL, TRPL and diffusion microscopies experimental setups. It also includes additional spectral, time-resolved PL and diffusion data at different temperatures. Finally, it includes additional details on the drift-diffusion model at low temperature as well as output of this model for different physical parameters.

AUTHOR INFORMATION

**Corresponding Author**

*Samuel D Stranks

Cavendish Laboratory, University of Cambridge, JJ Thomson Avenue, Cambridge CB3 0HE, UK. *E-mail: sds65@cam.ac.uk

Department of Chemical Engineering & Biotechnology, University of Cambridge, Philippa Fawcett Drive, Cambridge CB3 0AS, UK

**Present Addresses**

†Géraud DELPORT has moved to: CNRS, UMR IPVF 9006, 18 boulevard Thomas Gobert, 91120 Palaiseau France

**Author Contributions**

The manuscript was written through contributions of all authors. All authors have given approval to the final version of the manuscript.

§ These authors contributed equally.

ACKNOWLEDGMENT

The authors acknowledge the European Research Council (ERC) under the European Union's Horizon 2020 research and innovation program (HYPERION, Grant Agreement Number 756962).

SDS acknowledges funding from the Royal Society and Tata Group (UF150033). GD acknowledges the Royal Society for funding through a Newton International Fellowship. GD and SDS acknowledge the UK Engineering and Physical Sciences Research Council (EPSRC) under grant reference EP/R023980/1. A.B. acknowledges a Robert Gardiner Scholarship and funding from Christ's College, Cambridge. K.G. acknowledges support from the Polish Ministry of Science and Higher Education within the Mobilnosc Plus program (GrantNo. 1603/MOB/V/2017/0). The authors thank Niall Goulding and Rachel Bothwell for valuable discussions.

# Supporting information:

## I. Synthesis of the crystals and samples preparation:

**Synthesis of (C₄H₉NH₃)₂(CH₃NH₃)₃Pb₄I₁₃ (BA₂MA₃Pb₄I₁₃; n = 4) bulk single crystal.** A temperature-programmed crystallization method was applied to synthesize n = 4 RPP single crystal[17]. PbO (0.69 M), BAI (0.17 M) and MAI (0.52 M) precursors were dispersed in a concentrated HI and $H_3PO_2$ pmixture (7.6:1, vol/vol) in an Ar-filled glove box, and then heated at 110 °C with stirring for 40 min to give a clear yellow solution. The solution was quickly transferred to an oven at 110 °C and allowed to cool slowly to room temperature at a rate of 3 °C h$^{-1}$, where upon metallic black square- or rectangle-shaped crystals started to form. The crystals were isolated by vacuum filtration and dried in an Ar-filled vacuum chamber at room temperature.

**Synthesis of (C₄H₉NH₃)₂(CH₃NH₃)₃Pb₄I₁₃ (BA₂MAPb₂I₇; n = 2) bulk single crystal.** PbO (0.59 M), BAI (0.43 M) and MAI (0.31 M) precursors were dissolved in a concentrated HI and $H_3PO_2$ mixture (9:1, vol/vol) in an Ar filled glove box. The subsequent steps are same as the preparation of n = 4 described above.

**RPP exfoliation**. Thin RPP flakes are produced by mechanical exfoliation of their bulk single crystals. Scotch tape is applied to bulk crystal surface before adhering to a substrate

## II. Determination of the number of RPP layers via optical contrast imaging:

We applied the optical contrast imaging method[1] (see Figure S1) to evaluate the thickness of the flake to be ≈9 stacked quantum wells (≈30 nm thickness). The smallest observed variation of the contrast ~10 arb. unit., is likely to correspond to the addition (or suppression) of a monolayer. Therefore, the thickness of flake of interest is estimated to be of 9 monolayers (~ 30 nm), given that its optical contrast is of ~ 90.

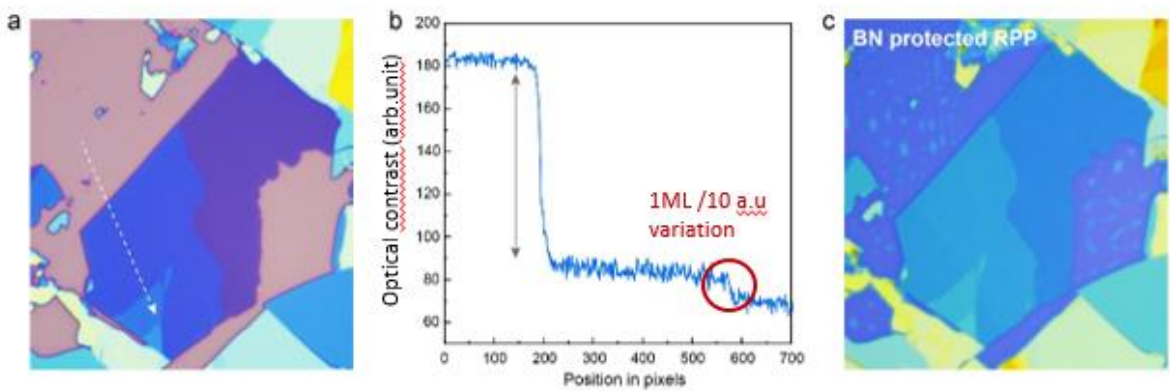

Figure S1 | **Determination of RPP layers by optical contrast imaging.** (**a**) Optical image of n = 4 RPP flakes exfoliated on $SiO_2$/Si substrate obtained in a reflection configuration. The different parts of the flakes exhibit different optical contrast (and colours). The optical contrast is a direct signature of the number of stacked quantum wells at a particular location of the sample. (**b**) Evolution of the optical contrast obtained along cross-section line on (**a**). (**c**) Optical image of BN protected *n* = 4 RPP flakes with the same area in (**a**).

### III. Details of the photoluminescence microscopy setup:

The time-resolved photoluminescence (TRPL) images and diffusion measurements were measured using a confocal microscope setup (PicoQuant, MicroTime 200, see Figure S3) The excitation laser, a 510-nm pulsed diode (PDL 828, PicoQuant, pulse width of around 100 ps), was directly focused onto the sample with an air objective. The emission signal was separated from the excitation light using a dichroic mirror. A pinhole of 50μm was included in the detection path, as well as an additional 510-nm longpass filter to minimise the laser contribution to the recorded signal. The TRPL was then focused onto a Hybrid PMT detector connected to a Picoquant acquisition card for time correlated single-photon counting (time resolution of 100 ps). Repetition rates of 10 MHz were used for the maps and the diffusion profiles, depending on the configuration.

### IV. Installation of a cryostat for the temperature dependent measurements

For the TRPL and diffusion mapping as a function of temperature, the perovskite crystals were placed under vacuum and mounted to the cold finger of a constant-flow liquid helium cryostat (Oxford, MicroHires).

### V. Confocal mapping and Diffusion mapping configurations

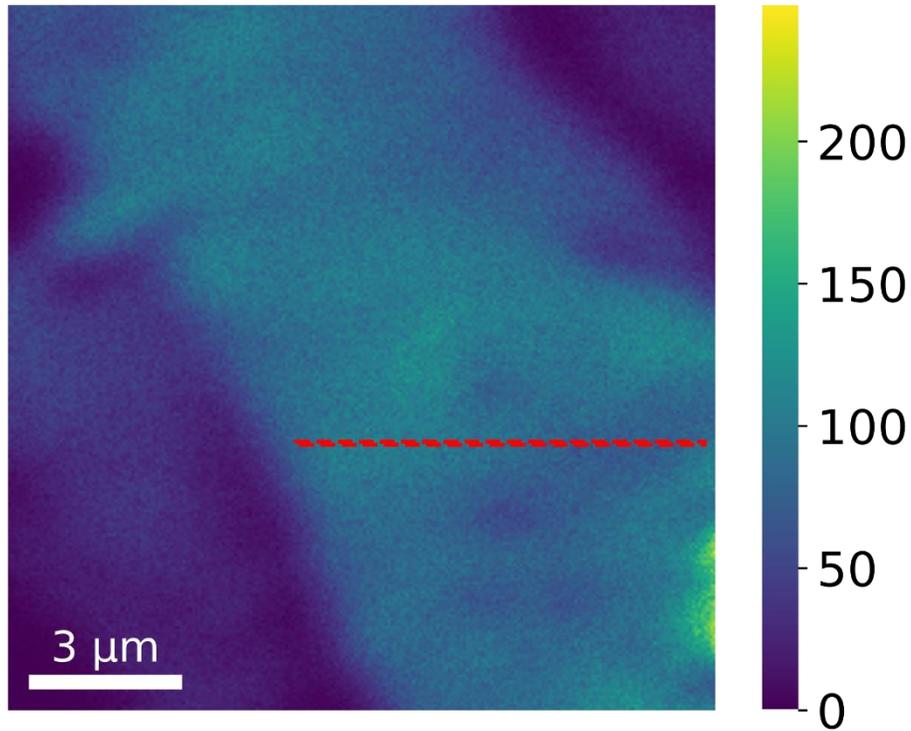

**Figure S2: PL intensity map of n=2 flake (excitation at 510 nm) at room temperature**. images. The red strip highlights the region of interest over which transport is subsequently analysed, with the spatial parameter x defined spanning from the centre of the region.

The raster scanning was performed using a galvo mirror system while both the objective and the sample remain at a fixed position. In the case of conventional local TRPL measurements with this setup, both the excitation and the emission are scanned through the mirror system. On the other hand, only the emission path was scanned to create the diffusion profiles from the main text, while the excitation was decoupled and fixed at the centre of the sample (x=0).

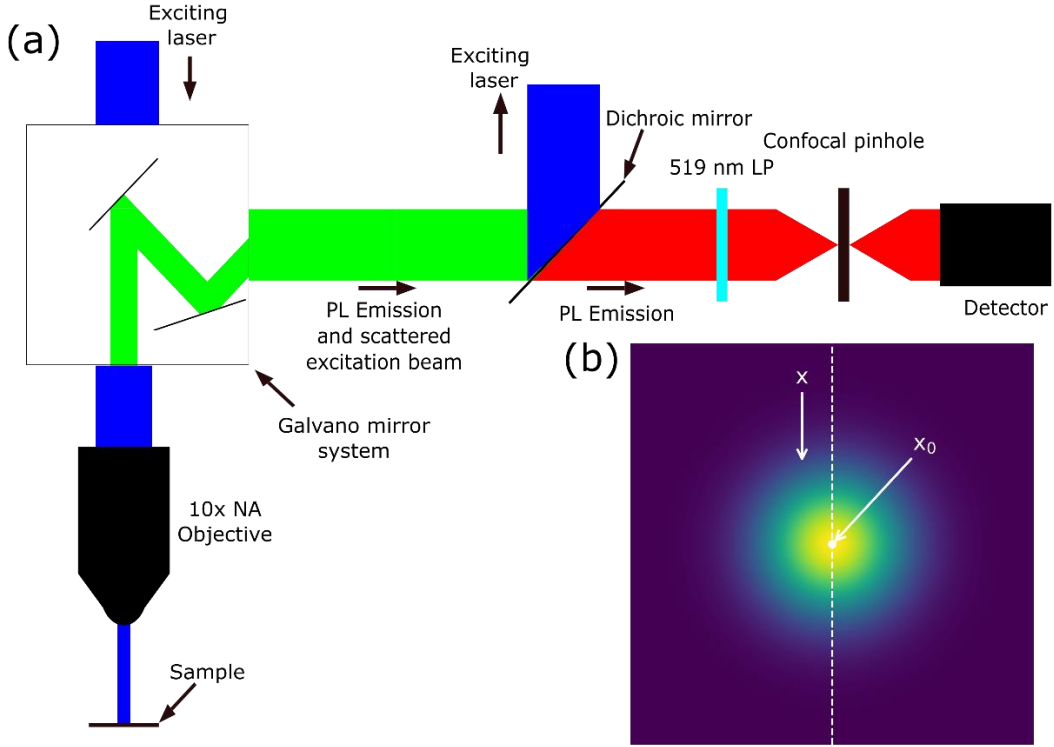

**Figure S3: Diffusion measurement technique.** (a) Confocal microscope configuration for diffusion measurements. Excitation laser show in blue, PL emission and scattered laser in green, PL after scattered laser removed in red. In contrast to the normal working mode the excitation source does not enter the galvano mirror system. (b) The sample excitation is fixed at x = 0, x0, while the PL emission, heatmap, is scanned along the x axis, dashed white line.

## VI. Used objective lenses and optical resolution

For the PL maps (Figure 1c and 2a) and the diffusion measurements displayed in (Figure 2b), a 100x air objective of 0.8 Numerical aperture (NA) was used. In this configuration, the lateral spatial resolution is of the order ~500 nm. As a demonstration of this, the standard deviation of the measured Gaussian beam at t = 0 in Figure 2a is of $\sigma(0)$ = 528 nm.

For the rest of the diffusion measurements and TRPL mentioned in the main text (Figure 2c, 3 and 4), a 10x objective lens with a 0.4 NA has been used. Indeed, we found such a low magnification objective has a long working distance that is more suitable for the cryogenic measurements. As seen in Figure S4, this objective lens populates the material with excitons over a larger spatial profile ($\sigma(0)$ = 2165 nm), but does not alter the diffusion measurements.

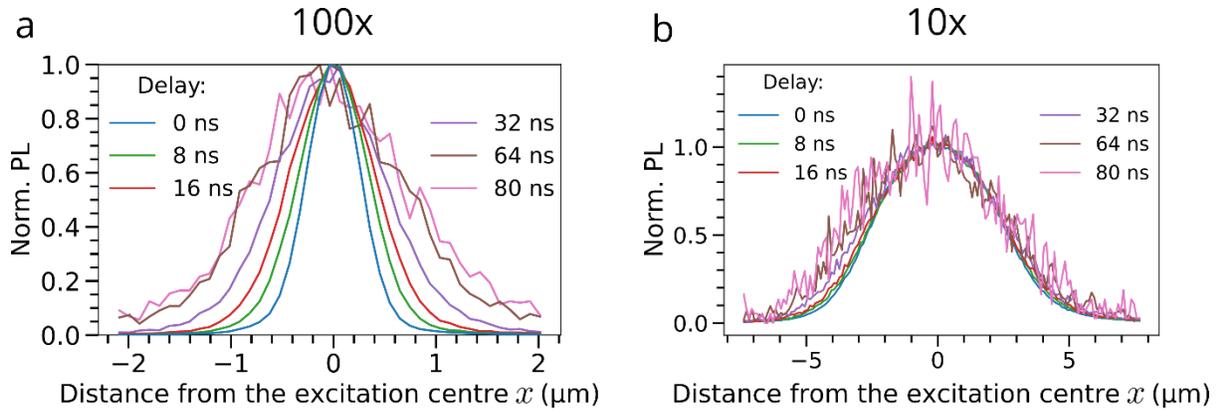

**Figure S4: Comparison of PL profiles for the objective lenses used.** (a) 100x 0.8 NA objective and (b) 10x 0.4 NA objective. (a) is the same as presented in Figure 2(b). Excitation was via a 510 nm 10 MHz laser.

## VII. Factors influencing the value of σ(0) valued obtained from the diffusion measurements.

The initial value of σ at t= 0 originates from a combination of factors, including the optical resolution of the setup (see discussion in section VI) and the possibility of early time diffusion or reabsorbed photons emitted at early times[2] within the temporal instrument response of the setup (≃100 ps).

## VIII. PL spectra measurements and hyperspectral mapping of the perovskite flake of interest.

Unless mentioned otherwise; all of the PL spectra included in this paper were measured directly inside our confocal PL microscope (including the temperature dependent PL spectra). To obtain these spectra, the collection signal of the microscope was diverted to ANDOR Kymera 193i spectrograph with a 600 lines per mm blazed at 500 nm, coupled to a CCD Array camera.

The only exception is the PL spectral maps of Figure S5, that were performed with a wide-field microscope (IMA VISTM, Photon Etc.). We employed a wide field air objective of 20x magnification (0.45 Numerical aperture (NA)) to characterise the flake in reflectance mode. The PL measurements were carried out using excitation from a 405 nm continuous wave laser, focused on the sample surface. The reflected laser beam was removed through use of a 420 nm long pass filter allowing collection of the emission spectrum of the sample. In order to obtain spectrally resolved images, a volume tunable Bragg filter was employed.

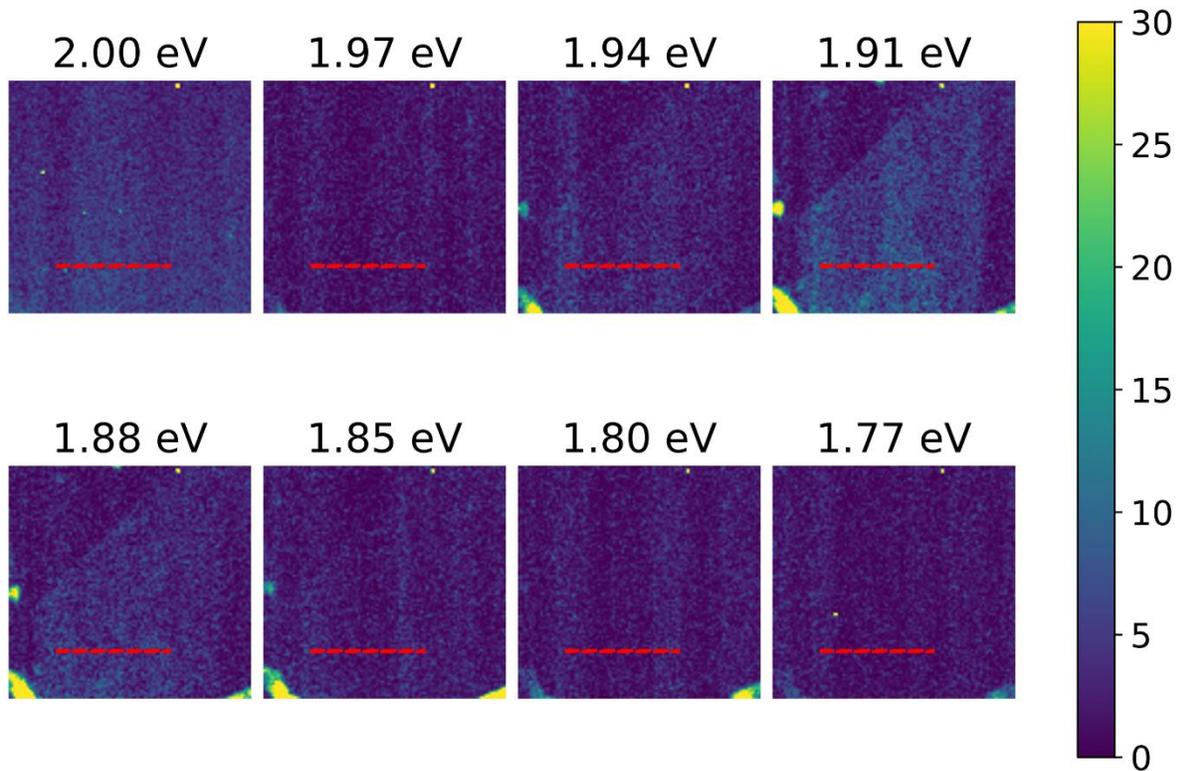

**Figure S5 Room temperature hyperspectral PL maps for the n = 4 flake**, with the investigated area shown in red. These images demonstrate that most of the PL signal originates from the main resonance at 1.9eV (see Figure S6 below). To increase the contrast, the 2 eV image, showing only noise, was used as an estimate of the background, with all other images subtracted having this subtracted having this subtracted. The 2 eV image is shown as recorded less 285 counts to ensure the common colour bar remains applicable for all images.

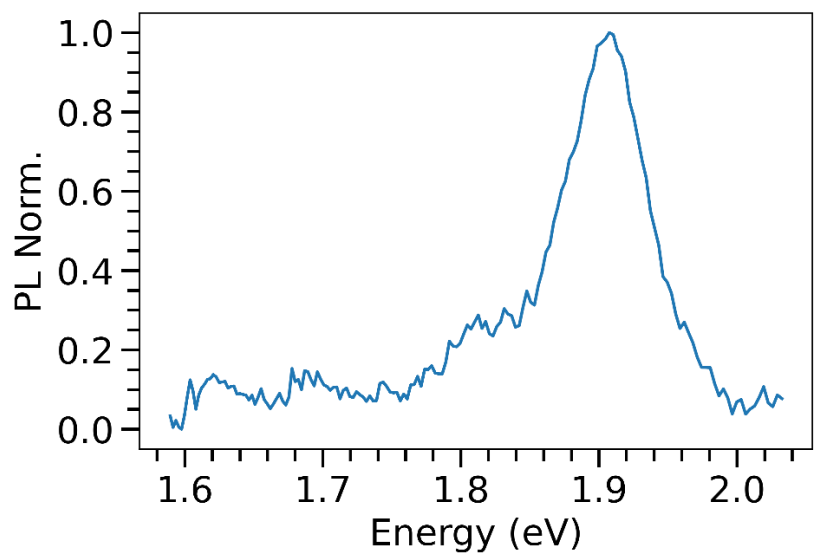

**Figure S6: Spatially integrated PL spectrum of the investigated n = 4 flake**, obtained from hyperspectral mapping in Figure S5 normalised for clarity. With a resonance around 1.9 eV, this spectrum corresponds to previously reported values[3] for n=4 crystals.

## IX. Additional TRPL data for the fluence series at room temperature

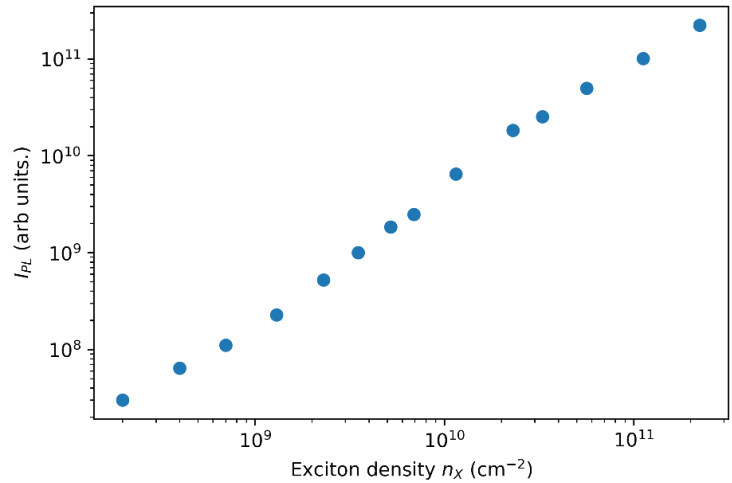

**Figure S7: Evolution of the PL intensity as a function of exciton density of the n=4 RPP flake,** from which the values of the parameter η displayed in Figure 1d are calculated using the formula η = $I_{PL}/n_X$. The superlinear evolution of the PL intensity with fluence, that occurs for excitons densities between $1.10^9$ and $2.10^{10}$ cm$^{-2}$ indicates that the light induced trap filling process is taking place.

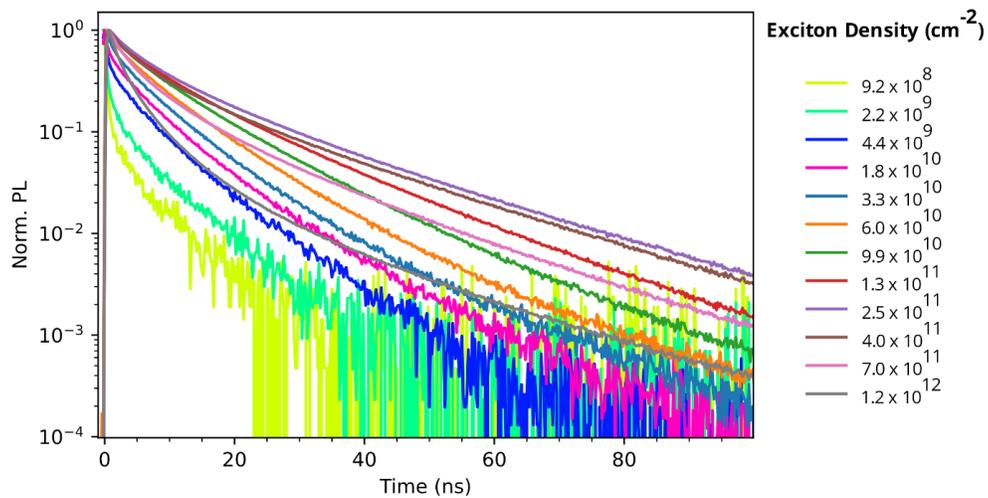

**Figure S8: Normalized TRPL decays of the n = 4 flake,** featuring more excitation densities than displayed in Figure 1c.

# X. Additional Diffusion data across for the fluence series at room temperature

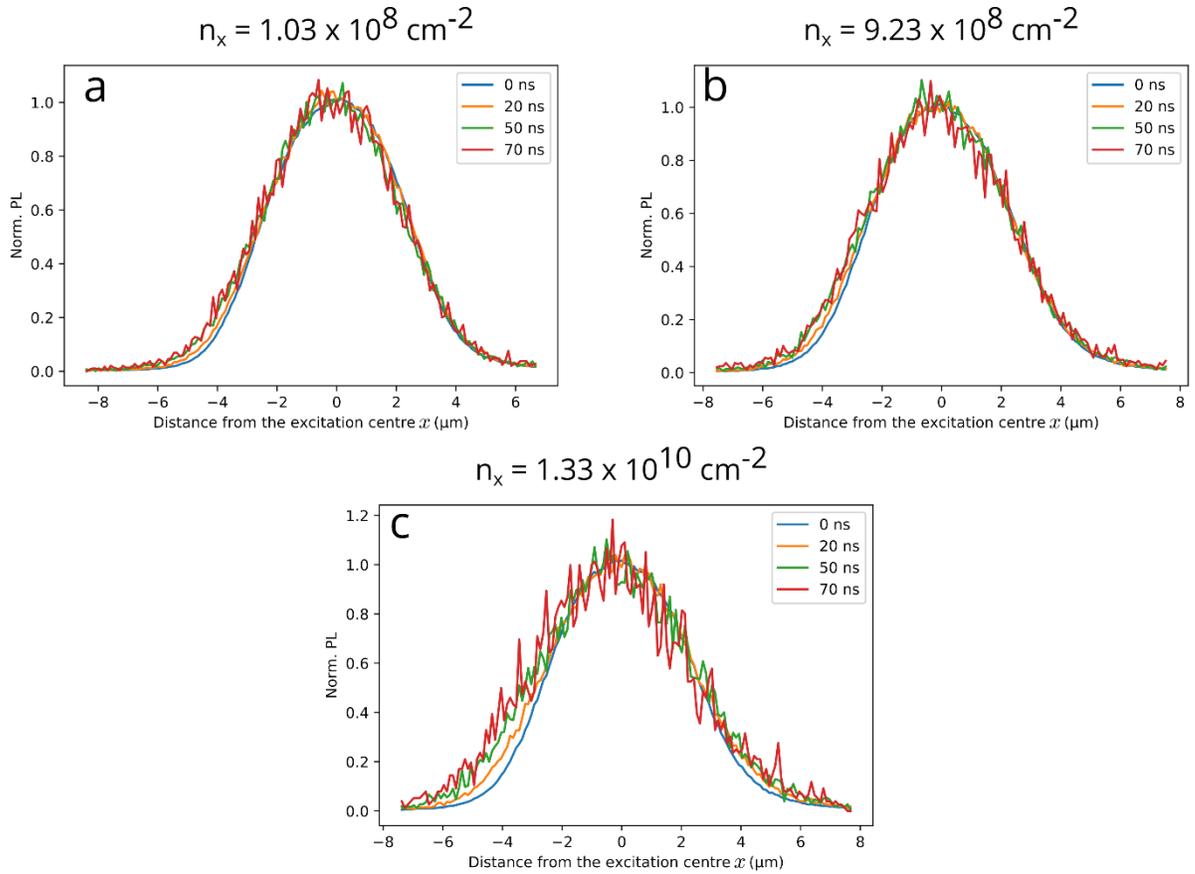

**Figure S9: Selected spatial PL profiles, normalised, at different times after excitation by the laser pulse,** localised at *x = 0* (centre of the red line in Figure 2a) from which the data in Figure 2c is obtained, taken with a 0.4 NA 10x objective lens. While the differences of the lateral width as a function of time seem small on these graphs, they are clearly quantifiable once we apply our Gaussian fitting process. The corresponding excitons densities $n_x$ are mentioned above each graph. The spreading of the spatial profiles becomes more pronounced as $n_{x,}$ as displayed in Figure 2c. Excitation was via a 510 nm 10 MHz laser.

# XI. Additional PL data as a function of temperature

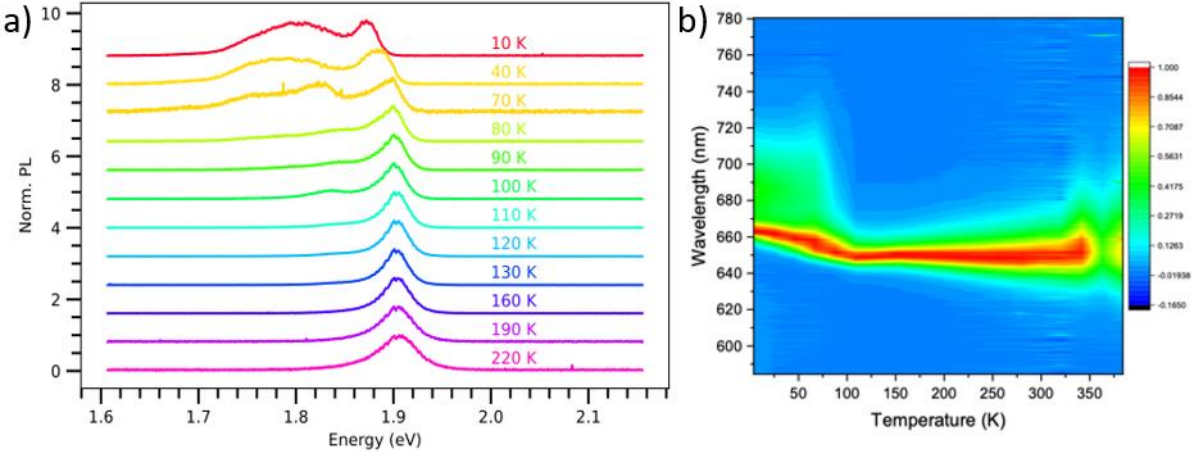

**Figure S10: Extended versions of Figure 3a, showing the Temperature dependent** evolution of the high energy (HE) PL peak and the emergence of the low energy (LE) peak present in the n=4 flake, with excitation at 510 nm and a repetition rate of 10 MHz. In a), the different spectra are normalised to the maximum value and vertically offset for clarity. In b), the spectra have been artificially collated into a contour plot, in which each colour corresponds to a different (normalized) PL intensity. Such contour plot is particularly suitable to highlight the threshold temperature of ~ 120-100 K under which the spectral position of the PL maximum starts changing.

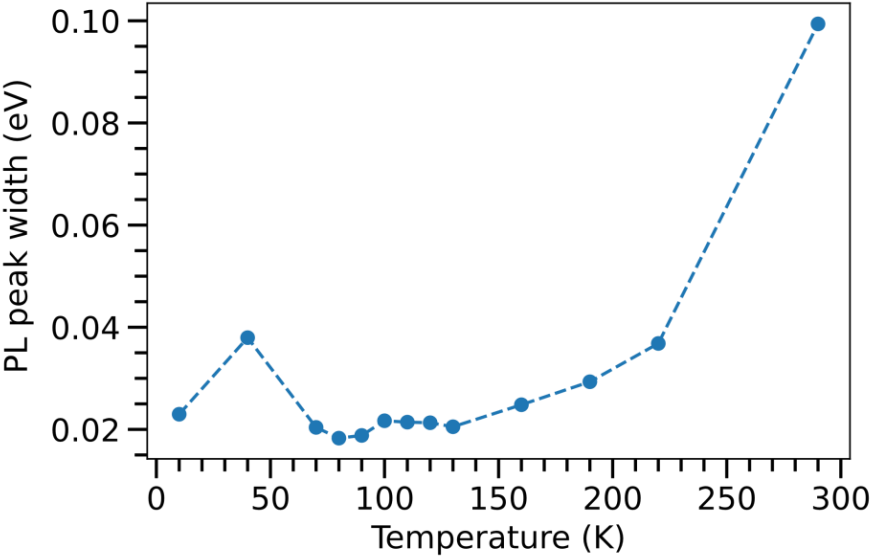

**Figure S11: FWHM of the HE resonance in the PL spectra of the *n=4* flake as a function of temperature,** corresponding to the spectra featured in Figures 3a and S10. The broadening the PL peak above 130 K is due the coupling of the excitons to the optical phonons that become available at these temparatures.

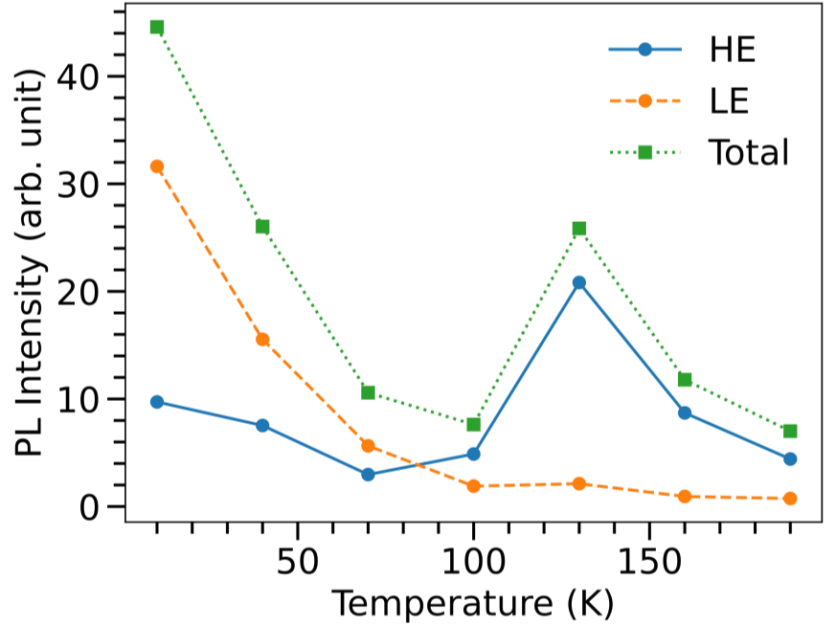

**Figure S12: Integrated PL intensity spectrally separated into the LE and high energy (HE) components,** shown with the total PL signal integral in green. Obtained from the PL spectra in Figure S10.

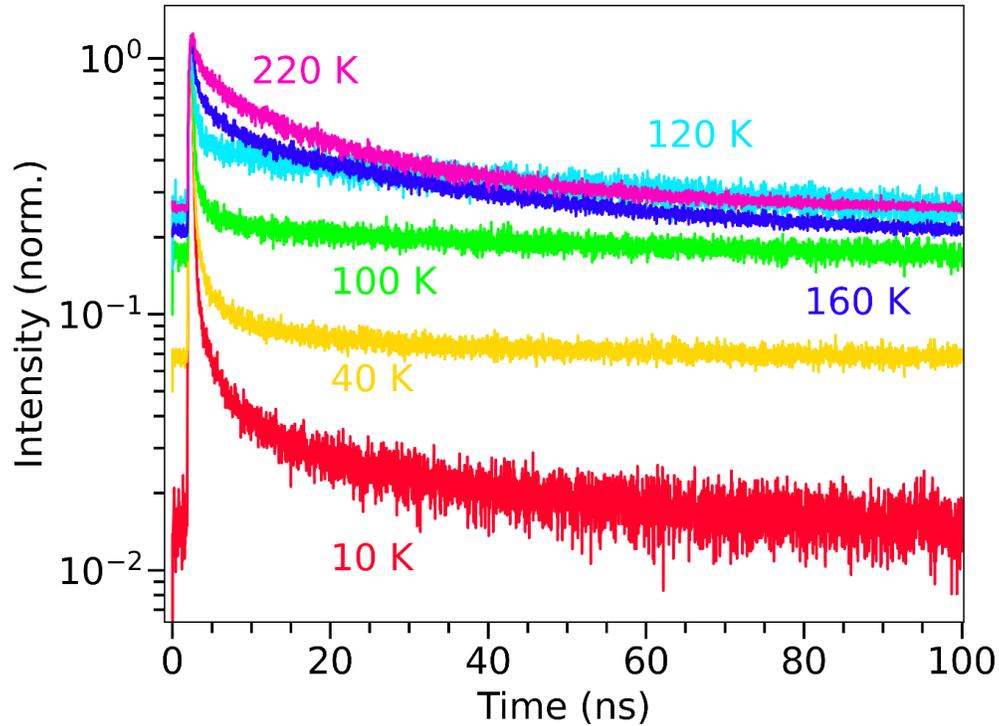

**Figure S13: TRPL decays of the n = 4 flake at different temperatures, o**btained from the centre of the region shown in Figure 1(c). As the temperature decreases, we observed a decrease of the slope at early time (< 5ns) of the TRPL decay curves, associated with the decrease of the 1/e

lifetime displayed in Figure 3b. In addition, we see the emergence of a long decay component as the temperature decrease, which is discussed in the main text.

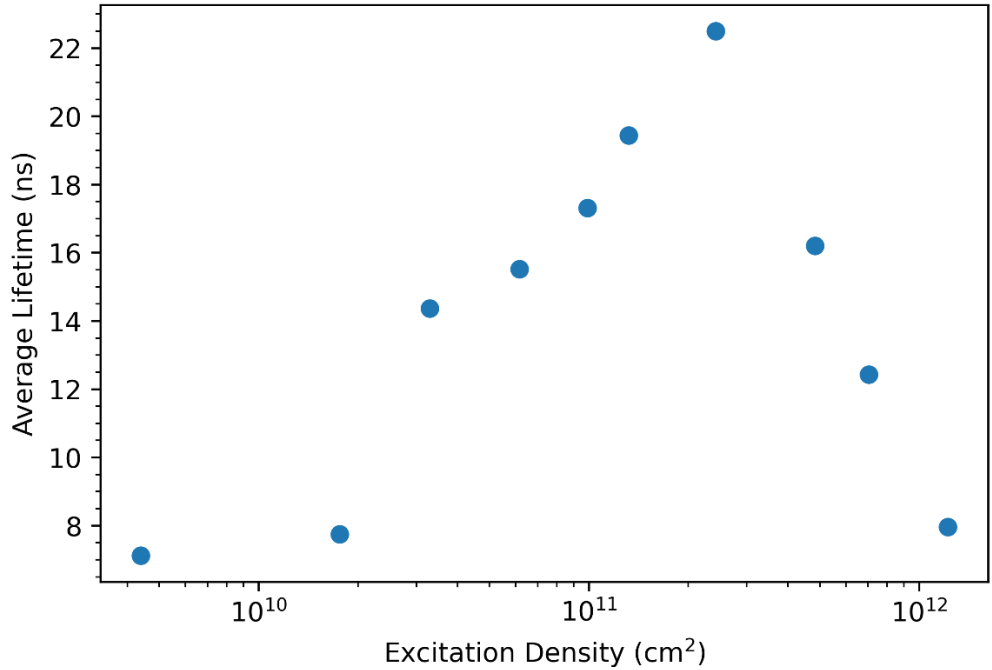

**Figure S14: 1/e lifetime of the n=4 flake at room temperature as a function of excitation density.** This is an extended version of the inset in Figure 1(e).

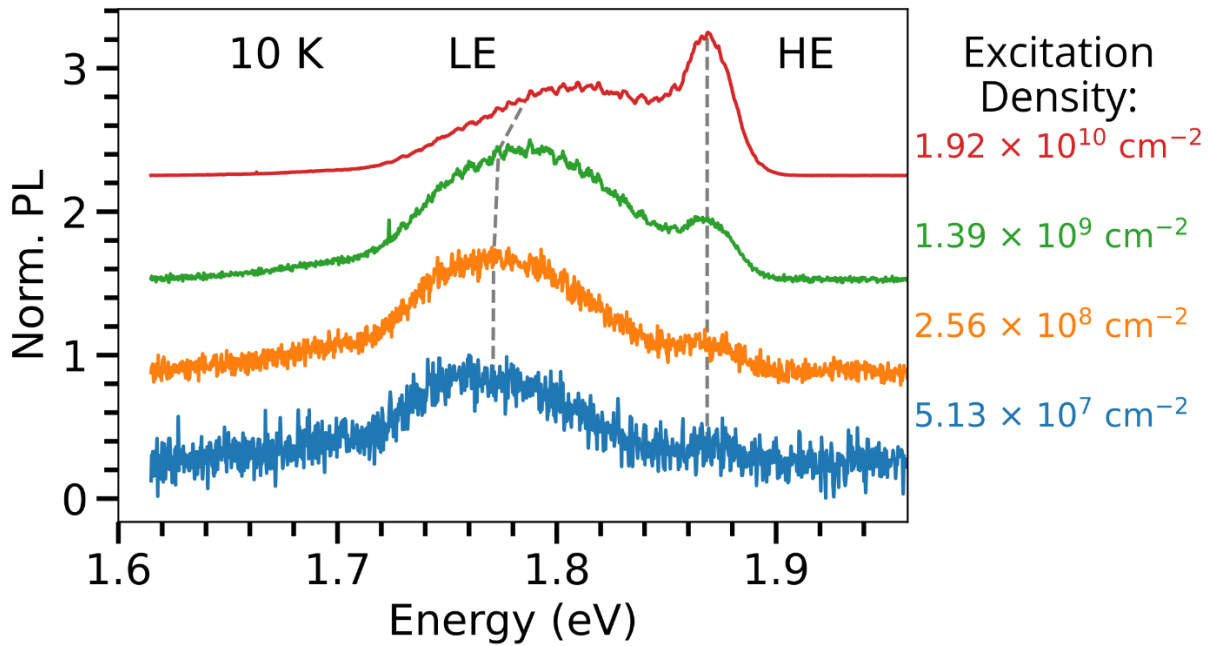

**Figure S15: PL spectra of the centre of the region of interest of the n=4 flake obtained at 10 K at difference excitation densities.** Saturation of the LE component with increasing excitation density can be seen to occur. Excitation was via a 510 nm laser at a repetition rate of 10 MHz.

# XII. Additional Diffusion profiles and extracted data

$$\sigma^2(t) = \sigma^2(0) + 2Dt \quad (1)$$

Equation 1 reproduced from the main text.

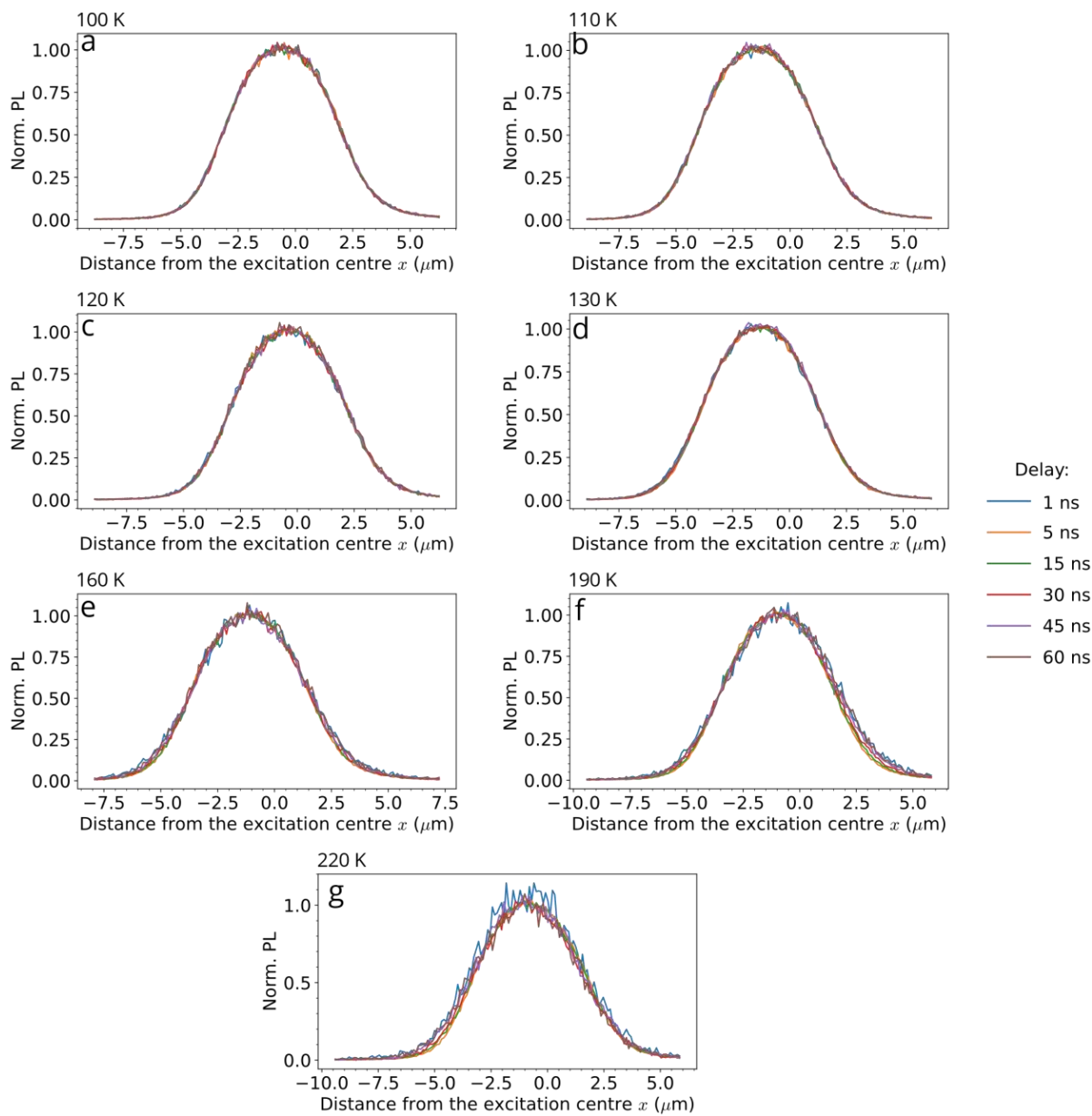

**Figure S15: Selected spatial PL profiles for temperatures in the range 100 – 220 K, normalised, at different times after excitation by the laser pulse,** localised at x = 0 (centre of the red line in Figure 2a) taken with a 0.4 NA 10x objective lens. While the differences of the lateral width as a function of time seem small on these graphs, they are clearly quantifiable once we apply our Gaussian fitting process.

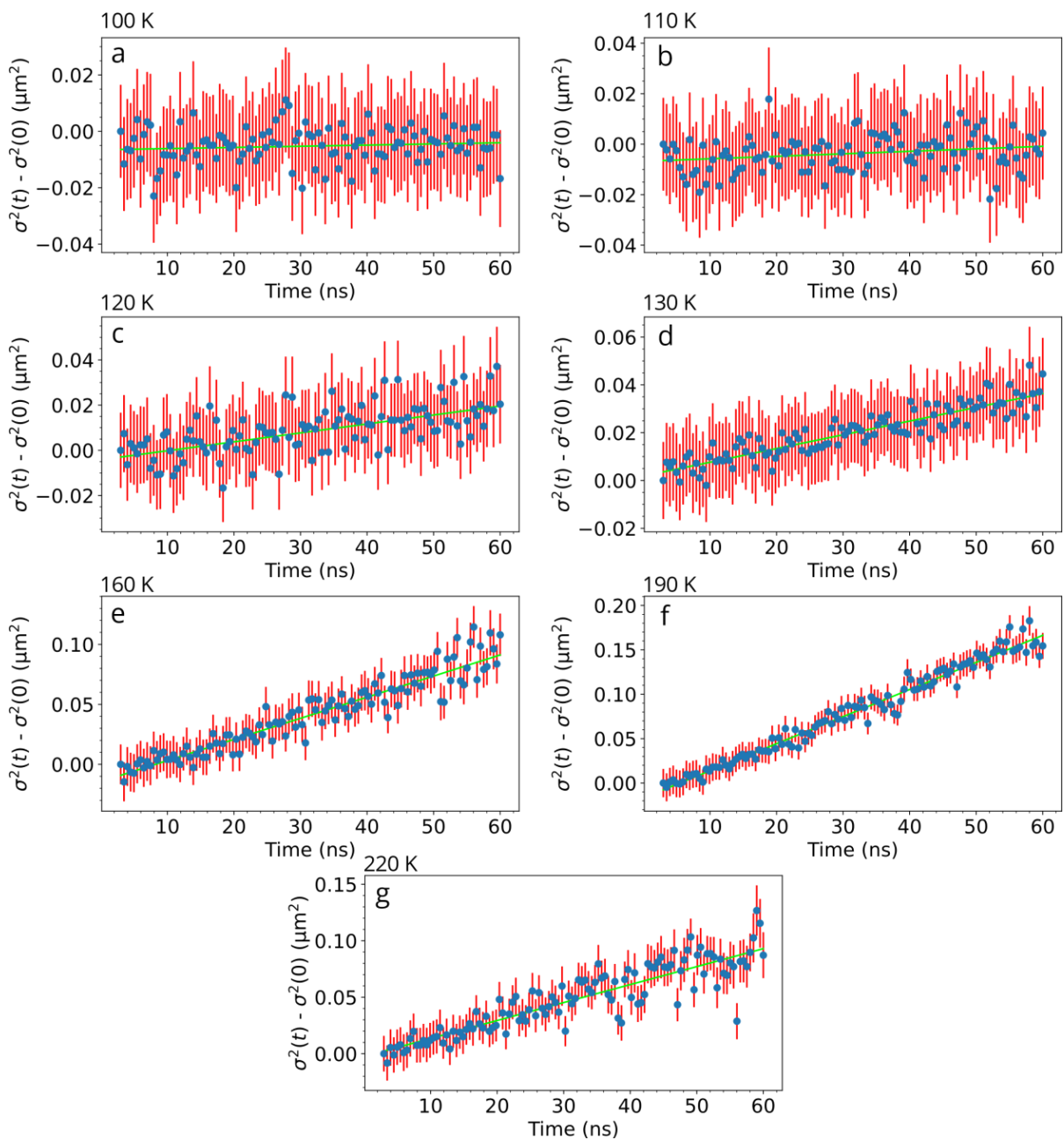

**Figure S16: Spreading profiles for temperatures in the range 100 K – 220 K obtained from the n = 4 flake.** Data in Figure 4c have been obtained from the fits to Equation 1 shown, green, in the above. These figures show a quasi-linear evolution of $\sigma^2$ with time which is consistent with a diffusive motion of excitons at these temperatures.

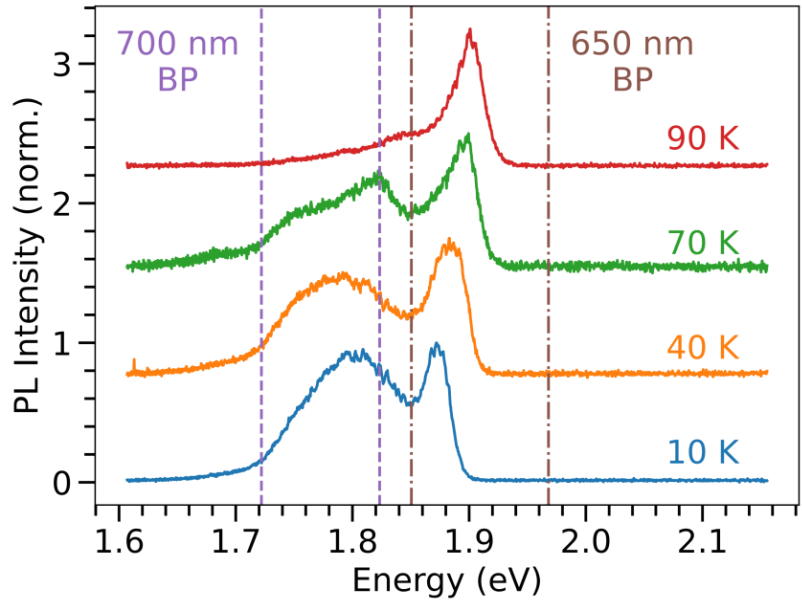

**Figure S17: PL spectra at the temperatures band pass filters were used**, the limits of the transparency windows of the two band pass filter ranges are indicated by vertical purple and brown lines.

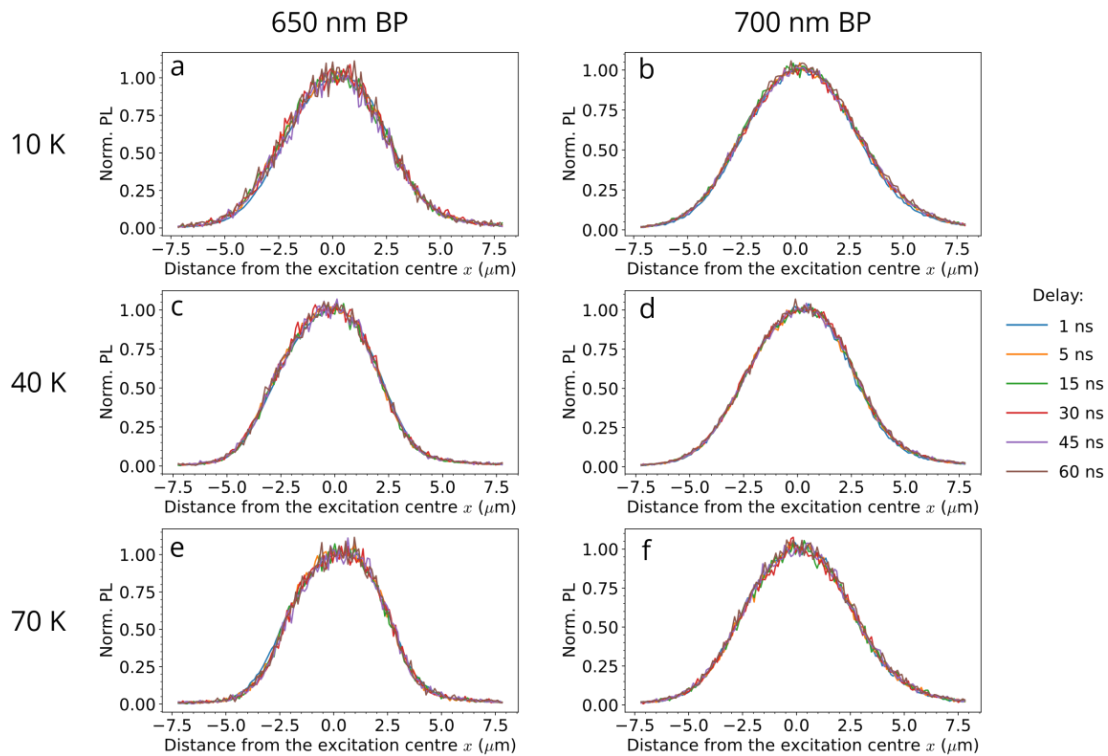

**Figure S18: Selected spatial PL profiles for temperatures in the range 10 – 70 K, normalised, at different times after excitation by the laser pulse,** localised at x = 0 (centre of the red line in Figure 2a) taken with a 0.4 NA 10x objective lens.

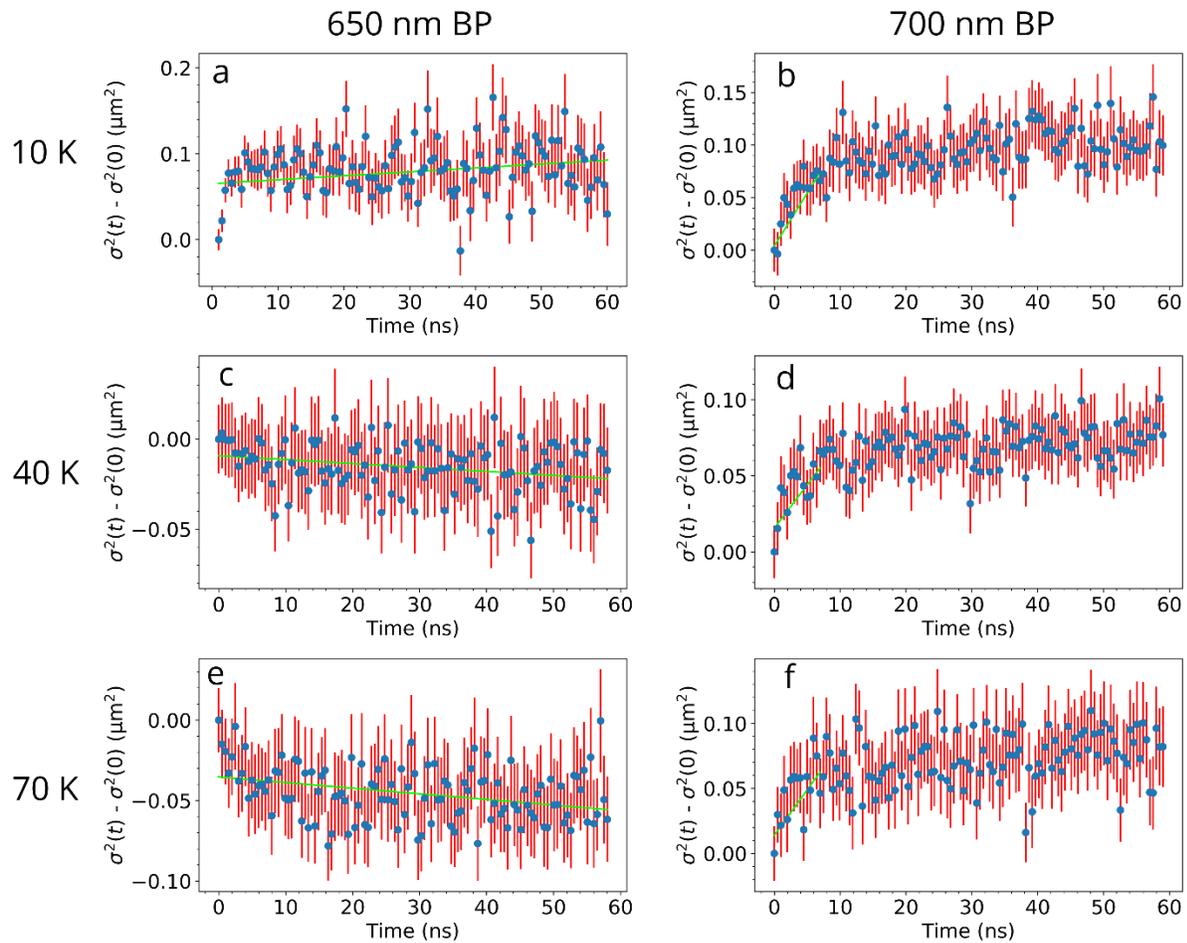

**Figure S19: Spreading profiles for temperatures in the range 10 K – 70 K obtained from the n = 4 flake.** Band pass filters have been used to enable the PL emission separated into the LE and HE components. Data in Figure 4c have been obtained from the fits shown, green, in the above. Figures b,d and f show a fast increase of σ² with time, followed by a saturation . As discussed in the main text, this effect is attributed to the efficient drift of excitons at early time, at these temperatures.

# XIII. Band Pass Resolved TRPL Decays at Low Temperatures

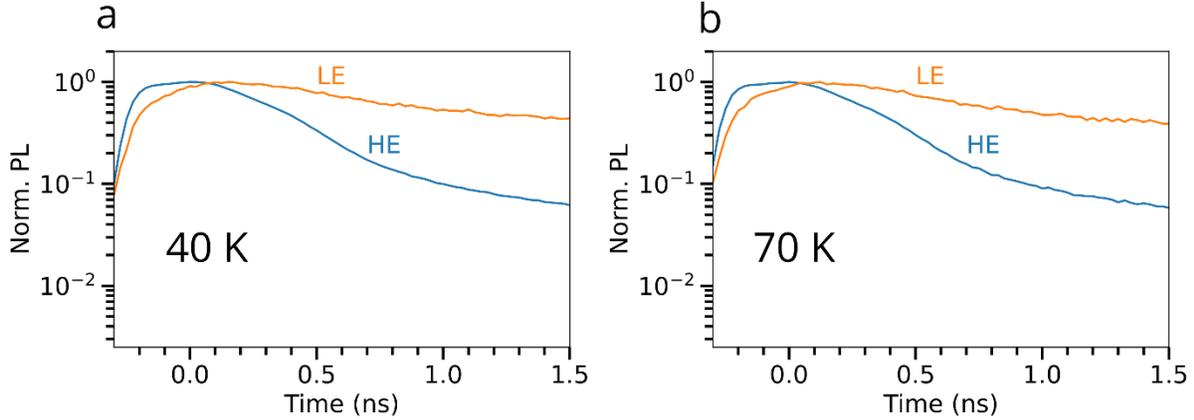

**Figure S20: BP resolved TRPL decays** at (a) 40 K and (b) 70 K obtained from the centre of the studied region of the n = 4 flake, highlighting the transfer from the rise time of the LE regions compared to the HE regions.

# XIV. Drift-Diffusion Model

$$\frac{\delta n(\vec{r},t)}{\delta t} = -kn(\vec{r},t) + D\nabla^2 n(\vec{r},t) + \mu\nabla\big(n(\vec{r},t)\nabla u(\vec{r})\big) \qquad (2),$$

Equation 2 reproduced from the main text.

**Model overview:**

The model used is based on Equation 2 and experimental data obtained at 10 K. Firstly BP resolved PL confocal images of the flake were obtained. These were individually normalised before being divided by each other to obtain an image showing the distribution of HE and LE domains in the flake, Figure 4d. The PL spectrum obtained at 10 K was normalised to ensure that the integrated intensity of this spectrum equalled to 1. This normalised PL spectrum was subsequently used as a probability density function from which the modelled energetic distribution of the flake was obtained. HE areas in the flake were assigned energies in the range (1.85 – 1.91) eV from the probability density function, while LE areas were assigned energies in the range (1.60 - 1.85) eV. The resulting energetic distribution was used to produce an energetic landscape within the flake used by the model.

Model parameters:

- Δt = 0.0005 ns;

- $\mu$ = 1.78 cm² eV⁻¹ s⁻¹;

- D = 0.2495 cm² s⁻¹;

- $k_{LE}$ = .01 ns⁻¹;

- $k_{HE}$ = .6 ns⁻¹.

The above parameters are the ones use by the model to produce the data in Figure 5. $k_{LE}$ and $k_{HE}$ are the exponential decay rates for the low and high energy regions respectively obtained from the experimental TRPL data, Figure 5d. µ is the effective mobility for the system which was calculated to be the mobility that would produce an average transfer rate of .5 ns⁻¹, as seen in the TRPL decays of Figure 5d, from the modelled energetic distribution in the flake. D, the diffusion coefficient in the material, remained the only free parameter which was manually adjusted to achieve good qualitative agreement with the experimental data.

Figure S19 below displays the outputs from simulations of our proposed model, using the Equation 2 from the main text, for parameters other than those considered in the main text. In particular, panel a corresponds to the purely diffusive case, in which we retrieve the fact that σ² is proportional to time, while panel b corresponds to limit in which only drift exist, leading to a sublinear evolution of the σ², sometimes referred as the ballistic regime[4].

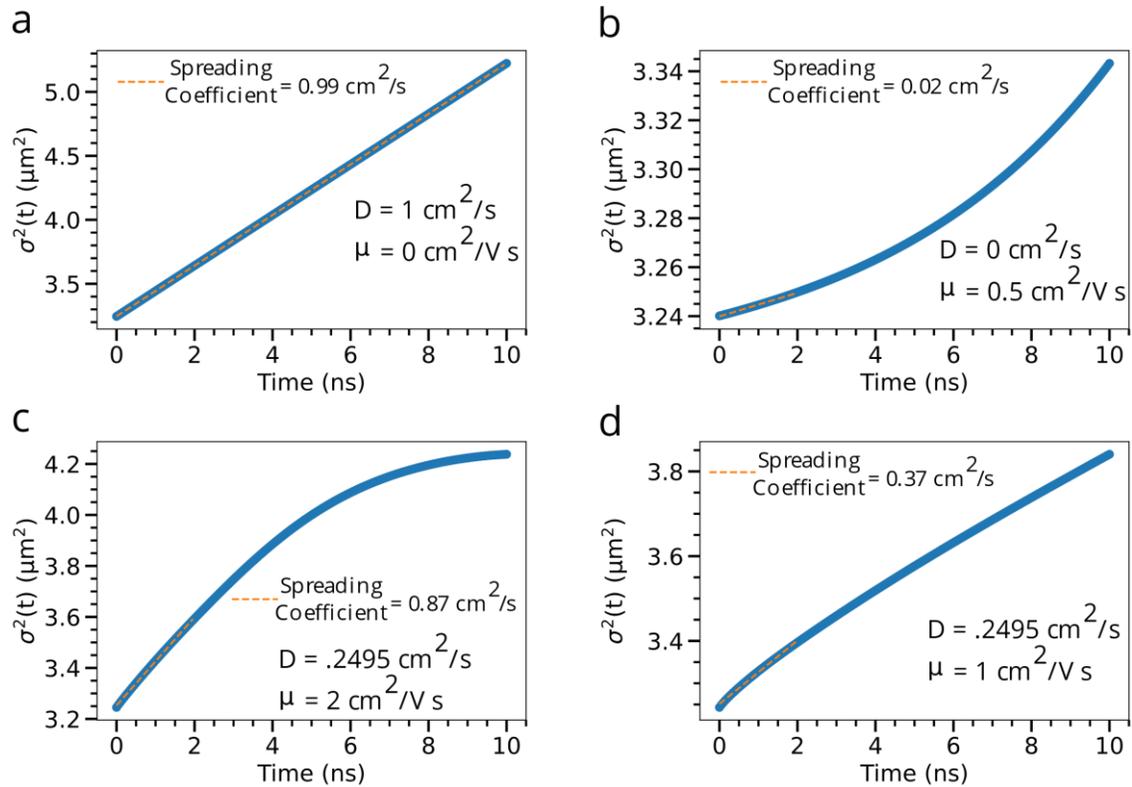

**Figure S21: Modelling spreading curves in an RRP flake at low temperature.** The used common parameters are detailed below, while any parameters which were altered are shown in the figure. Fits to the initial parts of the data are shown in orange for each resulting simulation.

## XV.   Complementary measurements for the n=2 flakes

These results are displayed to highlight the similarities between the excitonic behaviour within the *n*=4 flakes (in the main text) and the *n*=2 ones (below), despite the different in the size of the quantum wells.

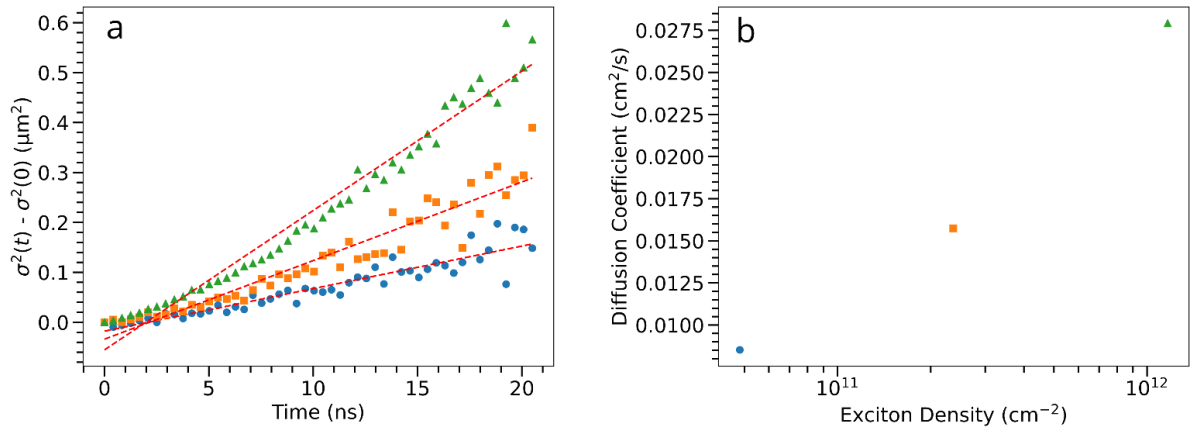

**Figure S22: Room temperature fluence measurements for n = 2 flake.** (a) Spreading profiles showing the temporal evolution of the squared broadening quantities $\sigma^2(t) - \sigma^2(0)$ of the spatial profiles for the flake at room temperature on the region of interest at three different densities of injected excitons. Fits to Equation 1 are shown in red. (b) Diffusion coefficients extracted from fits to the data with Equation 1 of the main text. As for the n=4, we see here that the measured diffusion coefficient increases with fluence as the trap filling process take places. This indicates that the presence of traps has an important influence on the excitons motion.

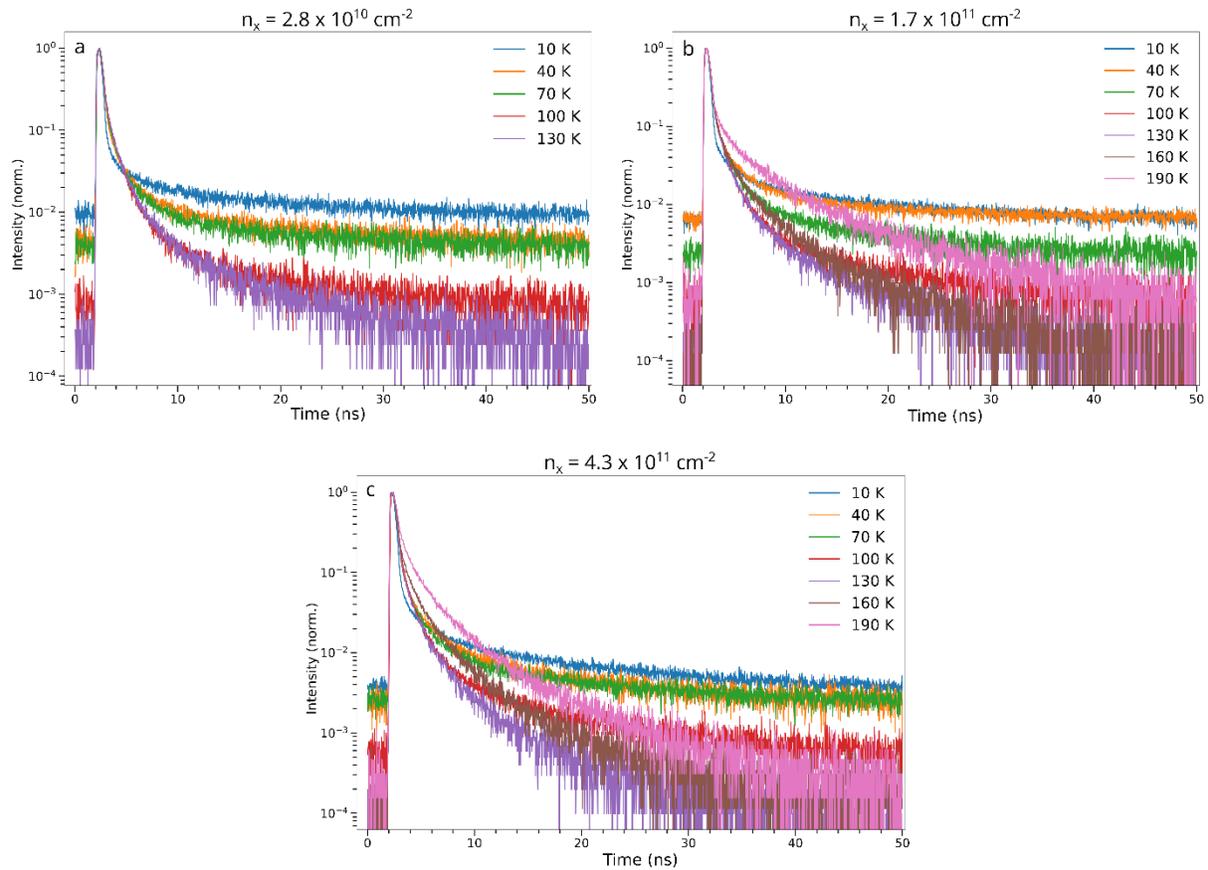

**Figure S23:** Time-resolved PL decays for the n = 2 flake shown in Figure S2 at different temperatures and excitation densities. Globally showing that the lifetime at early time decreases as the temperature decreases, similarly to the n=4 flakes.

## XVI. References.